\documentclass[usenatbib,useAMS]{mn2e}

\usepackage{natbib}
\usepackage{graphicx,epsfig}
\usepackage{latexsym,amssymb}
\usepackage[fleqn]{amsmath}

\newcommand{\hmsun}{\ensuremath{h^{-1}M_\odot}}
\newcommand{\msun}{\ensuremath{M_{\odot}}}

\newcommand{\hkpc}{$h^{-1}$\,kpc}
\newcommand{\hmpc}{$h^{-1}$\,Mpc}

\def\rhob{\bar{\rho}}
\newcommand{\beq}{\begin{equation}}
\newcommand{\eeq}{\end{equation}}
\newcommand{\beqa}{\begin{eqnarray}}
\newcommand{\eeqa}{\end{eqnarray}}
\newcommand{\rmd}{{\rm d}}

\newcommand{\dsr}{\ensuremath{\Delta\Sigma(R)}}
\newcommand{\ds}{\ensuremath{\Delta\Sigma}}
\newcommand{\avsigr}{\ensuremath{\overline{\Sigma}(<R)}}
\newcommand{\sigr}{\ensuremath{\Sigma(R)}}
\newcommand{\rmin}{\ensuremath{R_\mathrm{min}}}
\newcommand{\rmax}{\ensuremath{R_\mathrm{max}}}
\newcommand{\ups}{\ensuremath{\Upsilon(R;R_0)}}
\newcommand{\nt}{\ensuremath{N_{200}}}
\newcommand{\mtrue}{\ensuremath{M_{200b,\mathrm{true}}}}
\newcommand{\mest}{\ensuremath{M_{200b,\mathrm{est}}}}
\newcommand{\cvir}{\ensuremath{c_{200b}}}
\newcommand{\mvir}{\ensuremath{M_{200b}}}
\newcommand{\rvir}{\ensuremath{r_{200b}}}
\newcommand{\mvirt}{\ensuremath{M_{200b,20}}}
\newcommand{\zetac}{\ensuremath{\zeta_c}}
\newcommand{\mproj}{\ensuremath{M_\mathrm{2d}}}
\newcommand{\menc}{\ensuremath{M_\mathrm{3d}}}
\newcommand{\omegam}{\ensuremath{\Omega_m}}

\title[Lensing cluster masses]{Precision cluster mass determination from weak lensing}

\author[Mandelbaum et al.]{%
  Rachel Mandelbaum$^1$\thanks{\texttt{rmandelb@astro.princeton.edu}},
  Uro\v s Seljak$^{2,3}$, 
  Tobias Baldauf$^2$, 
  Robert E. Smith$^2$ 
  \\
  $^1$Department of Astrophysical Sciences, Princeton University, Peyton Hall, Princeton, NJ 08544, USA 
  \\
  $^2$Institute for Theoretical Physics, University of Zurich, Zurich Switzerland
  \\
  $^3$Department of Physics, University of California, Berkeley, CA 94720, USA
}

\begin{document}

\date{\today}

\maketitle 

\begin{abstract}
  Weak gravitational lensing has been used extensively in the past
  decade to constrain the masses of galaxy clusters, and is the most
  promising observational technique for providing the mass calibration
  necessary for precision cosmology with clusters.  There are several
  challenges in estimating cluster masses, particularly (a) the
  sensitivity to astrophysical effects and observational systematics
  that modify the signal relative to the theoretical expectations, and
  (b) biases that can arise due to assumptions in the mass estimation
  method, such as the assumed radial profile of the cluster.  All of
  these challenges are more problematic in the inner regions of the
  cluster, suggesting that their influence would ideally be suppressed
  for the purpose of mass estimation. However, at any given radius the
  differential surface density measured by lensing is sensitive to all
  mass within that radius, and the corrupted signal from the inner
  parts is spread out to all scales.  We develop a new statistic \ups\
  that is ideal for estimation of cluster masses because it completely
  eliminates mass contributions below a chosen scale (which we suggest
  should be about 20 per cent of the virial radius), and thus reduces
  sensitivity to systematic and astrophysical effects.  We use
  simulated and analytical profiles including shape noise to quantify
  systematic biases on the estimated masses for several standard
  methods of mass estimation, finding that these can lead to
  significant mass biases that range from ten to over fifty per
  cent.  The mass uncertainties when using the new statistic \ups\ 
  are reduced by up to a factor of ten relative to the standard
  methods, while only moderately increasing the statistical errors.
  This new method of mass estimation will enable a higher level of precision in
  future science work with weak lensing mass estimates for galaxy
  clusters. 
\end{abstract}


\section{Introduction}
\label{S:introduction}

Many scientific applications require robust measurements of the mass
in galaxy clusters.  One such application is the use of the dark
matter halo mass function to constrain cosmological model parameters,
including the amplitude of matter density perturbations, the average
matter density, and even the equation of state of dark energy
\citep[e.g., most recently,
][]{2007ApJ...657..183R,2008MNRAS.387.1179M,2009ApJ...692.1060V,2010ApJ...708..645R}.
Another example is validation and refinement of models of cluster
formation and evolution, which predict relations between the more
easily measured optical and X-ray emission, and the underlying dark
matter halo
\citep{2006ApJ...650..128K,2007ApJ...668....1N,2008A&A...482..451Z,2009arXiv0906.4370B}.
Currently, there are thousands of known clusters selected in various
ways that can be used for these applications.  Future surveys such as
the Dark Energy Survey
(DES)\footnote{\texttt{http://www.darkenergysurvey.org/}},
Pan-STARRS\footnote{\texttt{http://pan-starrs.ifa.hawaii.edu/public/}},
and the Large Synoptic Survey Telescope
(LSST)\footnote{\texttt{http://www.lsst.org/lsst}} will provide even
larger and deeper samples that can be used for this purpose, requiring
greater systematic robustness in the mass measures to complement the
smaller statistical errors.

Many different methods have been used to measure the halo profile of
clusters and thereby estimate their masses.  Kinematic tracers such as
satellite galaxies, in combination with a Jeans analysis or caustics
analysis, can give information over a wide range of physical scales
and halo masses.  While the issues of relaxation, velocity bias,
anisotropy of the orbits and interlopers need to be carefully
addressed, recent results suggest a good agreement with theoretical
predictions for the form of the density profile
\citep{2003ApJ...585..205B,2004ApJ...600..657K,2003AJ....126.2152R,2005ApJ...628L..97D,2006AJ....132.1275R,2007MNRAS.378...41S}.
Hydrostatic analyses of X-ray intensity profiles of
clusters 
use X-ray intensity and temperature as a function of radius to
reconstruct the density profile and estimate a halo mass.  The
advantage of thermal gas pressure being isotropic in partially lost 
due to the possible presence of other sources of pressure support,
such as turbulence, cosmic rays or magnetic fields. These extra
sources of pressure support cannot be strongly constrained for typical
clusters with present X-ray data \citep{2004A&A...426..387S}, but
could modify the hydrostatic equilibrium and affect the conclusions of
such analyses. Recent results are encouraging and are in a broad
agreement with predictions, although most require concentrations that
are higher than those predicted by a concordance cosmology
\citep{2006ApJ...640..691V,2007ApJ...664..123B,2007MNRAS.379..209S}.
While the above-mentioned systematic biases cannot be excluded, the
small discrepancy could also be due to baryonic effects in the central
regions, due to selection of relaxed clusters that may be more
concentrated than average \citep{2006ApJ...640..691V}, or due to the
fact that at a given X-ray flux limit, the more concentrated clusters
near the limiting mass are more likely to be included in the sample
\citep{2007A&A...473..715F}.

Gravitational lensing is by definition sensitive to the total mass,
and is therefore one of the most promising methods to measure the mass
profile independent of the dynamical state of the clusters.  Many
previous weak lensing analyses have focused on individual clusters
(for example,
\citealt{2007MNRAS.379..317H,2007ApJ...667...26P,2009ApJ...702..603A,2009arXiv0903.1103O}).
Measuring the matter distribution of individual clusters allows a
comparison with the combined baryonic (light and gas) distribution on an individual
basis, and so can constrain models that relate the two, such as MOND
versus CDM \citep{2006ApJ...648L.109C}.  However, these measurements
can be quite noisy for individual clusters.  Stacking the signal from
many clusters can ameliorate this problem, since shape noise and the
signal due to correlated structures will be averaged out.  Such a
statistical approach is thus advantageous if one is to compare the
observations to theoretical predictions, which also average over a
large number of halos in simulations.  A final advantage of stacking
is that it allows for the lensing measurement of lower-mass halos,
where individual detection is impossible due to their lower shears
relative to more massive clusters.  Individual high signal-to-noise
cluster observations and those based on stacked analysis of many
clusters are thus complementary to each other at the high mass end,
with the stacked analysis drastically increasing the available
baseline in mass.

Extraction of cluster dark matter halo masses from the weak lensing
signal is subject to a number of uncertainties, which we discuss in
this paper in detail, including the ways that the uncertainties differ
for individual versus stacked cluster lensing analyses.  In brief,
these uncertainties are: (i) biased calibration of the lensing signal;
(ii) modification of the lensing profile in the inner cluster regions
due to accidental inclusion of cluster member galaxies in the source
sample, intrinsic alignments of those galaxies, non-weak shear,
magnification, baryonic effects that modify the initial cluster dark
matter halo density profile, and cluster centroiding errors; (iii)
contributions to the lensing signal from nonvirialized local
structures and large-scale structure (LSS).  Furthermore, parametric
modeling of the mass requires the assumption of a form for the dark
matter halo profile, which may differ from the intrinsic profile
and/or have poorly constrained parameters.  Non-parametric modeling,
while not subject to this weakness, results in projected masses that
must be converted to three-dimensional enclosed masses to be compared
against the theory predictions, all of which are currently phrased in
terms of 3d masses.  We quantify the degree to which this conversion
depends on assumptions about the density profile.  Generally, we show
the effects of many of these uncertainties on the estimated masses
from cluster weak lensing analyses, both in the stacked and individual
cases, using parametric and non-parametric mass modeling.

Effects that modify the cluster density profile in the inner regions
($\lesssim 0.5$\hmpc), are particularly problematic given that the
weak lensing signal \dsr\ is sensitive to the density profile not just
at a projected separation $R$, but also at all smaller separations.
We propose a modified statistic, denoted \ups, that removes the
dependence on the projected density between $R=0$ and $R=R_0$, with
$R_0$ chosen to avoid
scales with systematic uncertainties.  The decrease in systematic
errors that results from removing scales below $R_0$ comes at the
expense of somewhat increased statistical errors.  We explore the
optimal choice of $R_0$, and quantify the degree to which our use of
this new statistic to estimate cluster masses lessens systematic biases and increases statistical
errors.  Our tools for this investigation include simple, idealised
cluster density profiles; more complex and realistic density profiles
from $N$-body simulations; and finally, real cluster lensing data from
the Sloan Digital Sky Survey (SDSS, \citealt{2000AJ....120.1579Y})
that was previously analysed by \cite{2008JCAP...08..006M}.

We begin in Section~\ref{S:theory} with a discussion of the
theoretical aspects of cluster-galaxy weak lensing, including a
detailed discussion of the challenges of mass determination, and a
summary of typical approaches to parametric and non-parametric mass
estimation, with the introduction of a new statistic from which to
derive parametric mass
estimates.  In Section~\ref{S:simulations}, we describe the $N$-body
simulations that we use to provide sample cluster density profiles.
Section~\ref{S:data} has a description of the SDSS cluster lensing
data we use to test for some of the effects that we find using the
simulations.  Results for both the theoretical profiles and the real
data are presented in Section~\ref{S:results}.  We conclude with a
discussion of our findings and their implications in
Section~\ref{S:conclusions}.

\section{Theory}
\label{S:theory}

This section includes theoretical background related to cluster-galaxy
weak lensing, modeling of cluster masses using lensing, and the new
statistic that we propose is optimal for cluster mass estimation.

\subsection{Standard lensing formalism}

Cluster-galaxy weak lensing provides a simple way to probe the
connection between clusters and matter via their cross-correlation
function $\xi_\text{cl,m}(\vec{r})$, defined as
\beq
\xi_\text{cl,m}(\vec{r}) = \langle\delta_\text{cl}(\vec{x})\delta_\text{m}^*(\vec{x}+\vec{r})\rangle_{\vec{x}},
\eeq
where $\delta_\text{cl}$ and $\delta_\text{m}$ are overdensities of
clusters and matter, respectively ($\delta_\text{m} =
\rho_m/\overline{\rho}_m-1$).  
This cross-correlation can be related to the projected
surface density
\beq\label{E:sigmar} 
\Sigma(R) = \overline{\rho} \int
\left[1+\xi_\text{cl,m}\left(\sqrt{R^2 + \chi^2}\right)\right] \rmd\chi,
\eeq 
where $\overline{\rho}$ is the mean matter density, $R$ is the
transverse separation and $\chi$ the line-of-sight direction over which we
are projecting.  Here, we ignore the line-of-sight window function, which is hundreds
of mega-parsecs broad and not relevant at cluster scales. For this
paper, we are primarily interested in the contribution to the
cluster-matter cross-correlation from the cluster halo density profile
$\rho_\text{cl}$ itself,
rather than from other structures, and hence
\beq\label{E:sigmar2}
\Sigma(R) \approx \int_{-\infty}^{\infty} \rho_\text{cl}(r=\sqrt{\chi^2+R^2})\,\rmd\chi.
\eeq

The surface density is then related to the observable quantity for
lensing, called the differential surface density,
\beq\label{E:ds} 
\dsr = \gamma_t(R) \Sigma_c= \avsigr - \sigr, 
\eeq
where $\gamma_t$ is the tangential shear (a weak but coherent distortion in the shapes of background galaxies)
  and $\Sigma_c$ is a geometric factor, 
\beq\label{E:sigmacrit} 
\Sigma_c = \frac{c^2}{4\pi G}
\frac{D_S}{D_L D_{LS}(1+z_L)^2}.
\eeq
Here $D_L$, $D_S$, and $D_{LS}$ are (physical) angular diameter
distances to the lens, to the 
source, and between the lens and
source, respectively. 
In the second relation in Eq.~\eqref{E:ds}, $\avsigr$ is the
average value of the surface density within some radius $R$,
\beq\label{E:avsig}
\avsigr = \frac{2}{R^2} \int_{0}^{R} R' \,\Sigma(R') \,\rmd R'.
\eeq
The second equality of Eq.~\eqref{E:ds} is true in the weak lensing limit, for
a matter distribution that is axisymmetric along the line of sight
(which is naturally achieved by the procedure of stacking many
clusters to determine their average lensing signal), or in the
non-axisymmetric case, provided that $\Sigma$ is averaged azimuthally.
For individual cluster analyses, 
profiles can be fit either using average shears in annuli, or with
full, two-dimensional shear maps.  

 Unless otherwise noted, all computations assume a flat
$\Lambda$CDM universe with matter density relative to the critical density $\omegam=0.25$ and $\Omega_{\Lambda}=0.75$.
Distances quoted for transverse lens-source separation are comoving
(rather than physical) \hmpc, where the Hubble constant $H_0=100\,h$
km $\mathrm{s}^{-1}\,\mathrm{Mpc}^{-1}$. 
Likewise, the differential surface density \ds{} is computed in comoving
coordinates, Eq.~\eqref{E:sigmacrit}, and the factor of $(1+z_L)^{-2}$
arises due to our use of comoving coordinates.

\subsection{Theoretical challenges in cluster mass modeling}\label{SS:challenges-theory}

In this section, we discuss theoretical challenges in cluster
mass modeling.  By ``theoretical'' challenges, we refer to issues that
cause the underlying cluster density profile (surface density
$\Sigma$) to be unknown.  This uncertainty in $\Sigma$ at a given
scale $R$ is propagated to larger scales in \dsr\ because of its
dependence on \avsigr\ (Eqs.~\eqref{E:ds} and~\eqref{E:avsig}).


\subsubsection{Unknown density profile}

When attempting to extract three-dimensional enclosed masses from the
projected lensing data, the unknown density profile may lead to a
biased mass estimate. For example, even for the latest generation of
simulations, the concentration parameter (defined more precisely
below) of clusters remains somewhat uncertain, with differences at the
level of 20 per cent at the high mass end
\citep{2004A&A...416..853D,2007MNRAS.381.1450N,2009ApJ...707..354Z}. The
concentration parameter at a given mass is also affected by the
assumed cosmological model, especially the amplitude of perturbations.
For a given halo mass, the differences between the profiles increase
towards the inner parts of the cluster, and if only those scales are
used in parametric fits for mass estimation, this can result in a
significant error on the halo mass.  In this paper, we investigate
bias due to unknown cluster concentration extensively, including the
use of parametric mass estimators with assumptions about the form of
the profile, and the use of non-parametric projected mass estimates
that require an assumption about the profile to get a 3d enclosed
mass.

\subsubsection{Baryonic effects}

The effect of baryons on the cluster mass distribution is unclear, but
may be significant in the inner cluster regions
\citep{1986ApJ...301...27B,2004ApJ...616...16G,2007ApJ...658..710N,2008ApJ...672...19R,2008PhRvD..77d3507Z,2009arXiv0907.1102B}.
Baryon cooling not only brings significant mass
into the inner regions of the cluster, but may also redistribute the
dark matter out to much larger scales than the scale of baryon
cooling.  These works suggest that the effect of baryons is to change
the cluster matter profile in the inner regions in a way that roughly
mimics a change in the halo concentration; however, the extent of this
effect in reality, and the affected scales, is unknown.  

\subsubsection{Offsets from minimum of cluster potential}\label{SSS:offsets}

The cluster centre about which the lensing signal should be computed
can be determined using a variety of methods. The most reliable
approach is to use the peak in X-ray or Sunyaev-Zeldovich flux. For
optically-identified clusters, the usual method is to find the
brightest cluster galaxy (BCG). The offsets from the true cluster
centre arise due to two effects: (1) BCGs may be slightly perturbed
from the minimum of the cluster potential well by some real physical
effect, such as an infalling satellite, and (2) photometric redshift
errors and/or limitations in the cluster detection technique (when
detecting clusters using imaging data) may lead to the wrong galaxy
being chosen as the BCG.  This latter effect might occur, for example,
with red-sequence cluster finding algorithms, in cases of BCGs with
bluer colours (estimated to be $\sim 25$ per cent of the BCG
population in reality, \citealt{2008MNRAS.389.1637B}).  As was
discussed quantitatively in \cite{2007arXiv0709.1159J}, the effect of
BCG offsets on stacked cluster lensing data is to convolve the surface
density $\Sigma(R)$ with some BCG offset distribution, which tends to
suppress the lensing signal in the inner regions (similar,
qualitatively, to the effect of the previous two systematic issues we
have discussed).  Consequently, fitted cluster masses and
concentrations will be reduced due to these centroiding errors
\citep{2002MNRAS.335..311G,2006MNRAS.373.1159Y}.  Note that
  while cluster centroiding errors arise due to
  observational limitations, we classify them as a theoretical
  issue because of their impact on $\Sigma(R)$ which leaks to larger
  scales in $\ds(R)$.

Studies comparing the BCG position to the cluster centre
defined by either the X-ray intensity or by the average satellite velocity
have found that the typical displacement is about 2--3 per cent of the virial
radius when the BCG is properly identified
\citep{2005MNRAS.361.1203V,2007ApJ...660..221K,2007ApJ...660..239K,2008MNRAS.389.1637B}.
The last of these studies finds that for about 10 per cent of BCGs, the
displacement is $>10$ per cent of the virial radius. Another study
that includes red galaxy photometric errors (i.e., both causes of
offsets rather
than just the first) finds that the median displacement is 10 per cent of the
virial radius \citep{2009ApJ...697.1358H}. 

Because the real data we use as a test case uses the maxBCG lens
sample, we focus in more detail on the issue of BCG offsets for
that cluster catalogue.  The maxBCG group uses mock catalogues to estimate
the distribution of BCG offsets resulting from the use of their
algorithm \citep{2007arXiv0709.1159J}.  
The accuracy of the distribution they find is quite
sensitive to the details of how the simulations are populated with
galaxies.  In brief, their result includes a richness-dependent
fraction of misidentified BCGs (from 30 per cent at low richness to 20
per cent at high richness), and those that are misidentified have a
Gaussian distribution of projected separation from the true cluster
centre, with a scale radius of $0.42$ \hmpc.

A full discussion of how these results from mocks compare with
  observations can be found in \cite{2008JCAP...08..006M}.  To
  summarize, at high masses (more than a few $\times 10^{14}$\hmsun), a comparison with X-rays
  \citep{2007ApJ...660..239K,2009ApJ...697.1358H} suggests that the
  mocks may overestimate the fraction of offsets greater than 250
  \hkpc.  However, the true level of offsets for the majority of the
  cluster catalogue is poorly constrained from the real data.

\subsection{Observational challenges in cluster mass modeling}\label{SS:challenges-obs}

In this section, we discuss observational challenges in
  cluster mass modeling.  We define ``observational'' challenges as
  those that result in difficulty in properly measuring $\ds(R)$ for a
  given density profile $\Sigma(R)$.

\subsubsection{Lensing signal calibration}\label{SSS:calibration}

The cluster-galaxy lensing signal overall calibration is an important issue for
cluster mass estimates.  The signal may be miscalibrated due to shape
measurement systematics \citep[e.g.,
][]{2006MNRAS.368.1323H,2007MNRAS.376...13M,2008arXiv0802.1214B},
unknown lens and/or source redshift distributions \citep[e.g.,
][]{2005A&A...439..513K,2008MNRAS.386..781M}, and contamination of the
``source'' sample by stars.  The effect of miscalibration typically is
to multiply the signal on all scales by a single multiplicative
factor.  We will investigate the effect of changes in lensing signal
calibration on the estimated masses when fitting both parametrically
and non-parametrically.

\subsubsection{Signal dilution due to cluster member galaxies}\label{SSS:dilution}

In principle, in the absence of intrinsic alignments, contamination of
the source sample by cluster member galaxies will dilute the cluster
lensing signal, since the cluster member galaxies are not lensed.
Thus, they suppress the cluster lensing signal, with the strongest
effect towards the cluster centre where the member galaxies are most
numerous.  For stacked cluster lensing data, this effect may be
effectively removed by cross-correlating random points with the source
catalogue, and boosting the signal by the scale-dependent ratio of the
weighted number of sources around the real clusters to that around the
random points
\citep{2004MNRAS.353..529H,2004AJ....127.2544S,2005MNRAS.361.1287M}.

For measurements of individual cluster lenses, the best way around
this problem is to use some colour-based criterion that removes the
cluster member galaxies.  
Without multicolour imaging, contamination of the
lensing signal can be several tens of per cent on a few hundred \hkpc\
scales \citep{2005ApJ...619L.143B,2007ApJ...668..643L}, and even with
it, there may be residual dilution of the signal of approximately ten
per cent on those scales
\citep{2007MNRAS.379..317H,2009arXiv0903.1103O}.  This scale-dependent
suppression of the signal results in  underestimation of the 
cluster mass and concentration.

\subsubsection{Intrinsic alignments}

Intrinsic alignments of galaxy shapes with the local tidal field can
affect cluster lensing measurements when cluster member galaxies that
are treated as sources actually have some mean alignment of their
shapes radially towards the cluster centre.  This effect, which leads
to a suppression of the lensing signal that is worse at smaller
transverse separations, has been detected observationally in several
contexts
\citep{2006ApJ...644L..25A,2006MNRAS.367..611M,2007ApJ...662L..71F,2007MNRAS.381.1197H,2009arXiv0903.2264S}.
Its amplitude varies with cluster mass, member galaxy type, and
separation from the cluster centre.  

The best way to avoid this effect
is to remove cluster member galaxies from the source catalogue, but a perfect removal 
is often not possible, as described in Section~\ref{SSS:dilution} and references therein.  When
using a very large stacked sample, the amplitude of the effect may be
roughly estimated using the estimated shear from the sample of
galaxies that were chosen based on the colour-redshift relation to be
cluster member galaxies.  This test, however, is only possible with good
colour information for the source galaxies.  We defer a detailed
discussion of the effects of intrinsic alignments on weak lensing
cluster mass estimates to future work, but the sign is always to lower
the signal (and therefore mass) in a way that is worse at smaller
cluster-centric radius.

\subsubsection{Non-weak shear and magnification effects}

The measured weak lensing signal is not precisely the
tangential shear $\gamma_t$, but rather the reduced shear
$g=\gamma_t/(1-\kappa)$, where $\kappa=\Sigma/\Sigma_c$ is the convergence.  For a typical cluster density profile, the difference between $g$ and
$\gamma_t$ is of order unity at the critical radius where $\kappa=1$ (that depends on the 
redshift, but can be as large as $100$\hkpc) reducing to  
a few per cent out to transverse separations of $\sim
500$\hkpc, beyond which the assumption that $g\approx\gamma_t$ is quite
accurate.  This distinction may be explicitly taken into account using
parametric mass models \citep{2006MNRAS.372..758M}, but is typically
ignored in non-parametric mass estimation (though since those
estimates usually do not rely on the shear on small scales, this
neglect is not necessarily a problem).

A related effect is magnification, which alters the source galaxy
population by changing the measured fluxes
and sizes\footnote{The change in
  apparent size may not be important for typical photometric data, but
  weak lensing measurements require imposition of an apparent size cut on
  the galaxies to ensure that they are well-resolved relative to the
  point-spread function (PSF).} \citep{2005MNRAS.361.1287M,2009PhRvL.103e1301S}.  As a
result, their redshift distribution may change, and the number
density of sources near lens galaxies typically differs from that in
the field.  Furthermore, if a correction is made to 
the observed weak lensing signal for stacked clusters to account for the
dilution due to cluster member galaxies included in the source sample,
as suggested in Section~\ref{SSS:dilution}, then this correction 
must be carried out by using the observed source number
  densities relative to that around random points.  The boost factor is supposed to only correct for changes in
  source number density due to clustering (which introduces unlensed
  galaxies into the source sample).  Since the number
  density of lensed galaxies may legitimately be altered by magnification,
   magnification can lead to
  incorrect boost factors.  
This effect may be accounted for using parametric mass modeling,
provided that the properties of the source sample at the flux and
apparent size limits is reasonably well understood
\citep{2006MNRAS.372..758M}.

\subsection{Summary of the challenges and how we model them}

The challenges discussed in the previous two subsections
  result in three types of changes in the lensing signal.  One type of
  change is an elevation (suppression) of the lensing signal on small
  scales that changes sign at some value of transverse separation to
  become a suppression (elevation).  For example, this change may
  result from an unknown dark matter concentration and baryonic
  effects.  The second type of change is a uniform suppression or
  elevation of the lensing signal in the inner cluster regions, such
  that the lensing signal gradually reaches the expected value at and
  above some value of transverse separation.  This change may result
  from cluster centroiding errors, dilution of the lensing signal due
  to cluster member galaxies and/or intrinsic alignments, non-weak
  shear, and magnification-induced errors in the source redshift
  distribution and number density.  The exact functional forms for and
  magnitudes of these changes, and their characteristic scale radii,
  vary depending on the situation.  However, we will use two models,
  one for each type of change.  The final type of change in the
  lensing signal that we consider is a uniform calibration offset.

The profiles we use for our
test cases include pure NFW profiles, and the cluster lensing
signal observed in $N$-body simulations. We modify the concentrations of
these test profiles, 
apply a model for the effects of cluster centroiding errors based
on mock catalogs \citep{2007arXiv0709.1159J}, and rescale them all to mimic calibration
offsets.  However, we do not want to rely too much on our modeling of
these effects (as concentrations and a specific centroid error model)
being correct in detail.  
Thus, if cluster mass
determination is to be robust, we 
need estimators that are as
insensitive to these types of changes in  cluster profile as
possible. Note that a key feature of all three types of changes in
profile is that they affect the inner cluster regions.  This fact leads to the requirement that the small scale
information is suppressed, which will motivate a new statistic 
introduced in this paper.

In all cases, we use spherically-symmetric profiles, as is appropriate
for stacked cluster lensing analyses.  The observed lensing
profile is roughly equivalent to the spherical average of the
underlying triaxial density profiles of the dark matter halos, so that
the cluster masses can be recovered to few per cent accuracy with mass
estimation assuming spherical profiles
\citep{2005MNRAS.362.1451M,2009MNRAS.396..315C}.  For individual
cluster lensing estimates, however, there is an additional level of
complication due to the assumption of a spherical profile: individual
deviations in the form of the profile from the assumed form due to
mergers, substructure \citep{2001A&A...378..748K}, and deviations from
a spherical shape \citep{2004MNRAS.350.1038C,2007MNRAS.380..149C} can
cause tens of per cent uncertainties in cluster mass model parameters.
We do not attempt to estimate the uncertainties for individual cluster
lensing analyses due to these effects, relying instead on previous
work.

\subsection{Signal due to other mass}

The measured lensing signal is caused by the projected mass distribution
around the cluster, and consequently it includes some contributions
that are not part of the cluster halo, which will affect the mass
estimates.  In the case of stacked cluster lensing analyses, the
average over these contributions from all clusters in the stack
results in the so-called halo-halo term, which can be modelled simply
using the cluster-matter cross-power spectrum as in, e.g.,
\cite{2000MNRAS.318..203S} and \cite{2005MNRAS.362.1451M}.  This term
becomes dominant on several \hmpc\ scales.  While here we use scales
where this term is sub-dominant, we will consider the question of how
the estimated masses may be biased if this term is not explicitly
modelled but is instead neglected. This failure to model the halo-halo
term should tend to pull the mass estimates upwards, since mass that
is not part of the cluster mass distribution will be attributed to the
cluster.  Our approach is to simply use the cluster lensing signal
from simulations without explicitly decomposing it into one- and
halo-halo terms; thus, mass that is not part of the cluster mass
distribution itself is implicitly included in our numerical
predictions of the cluster lensing signal.

For individual cluster lensing analyses, the effect of matter that is
not part of the cluster on the lensing signal is more complex, because
unlike for stacked analyses, no averaging process occurs over the
structures around many clusters.  As a result, local nonvirialized
structure \citep{1999ApJ...520L...9M,2001ApJ...547..560M} and
large-scale structure
\citep{2001A&A...370..743H,2003MNRAS.339.1155H,2004PhRvD..70b3008D}
can appear in the cluster lensing signal on all scales, not just large
scales, causing both an average bias and significant scatter in the
mass estimates.  A recent numerical study of large-scale structure projection
effects on weak lensing cluster counts \citep{2009ApJ...698L..33M}, has
shown that, whilst there is scatter and bias in the $M_{\rm
2d}$--$M_{\rm 3d}$ relation, the utility for such data to constrain
cosmological parameters through the mass function is not
impaired. Moreover, if one uses carefully constructed aperture mass
shear filters, then the bias arising from `correlated' large-scale
structure can be reduced to the percent level (Marian et al. ApJ
submitted.). However, the impact of `chance' projections along the
line-of-sight on the mass estimates is still relatively poorly
quantified.  While we use simulations to assess the effect of the
halo-halo term on stacked cluster analyses that neglect it, a detailed
treatment of this issue for individual cluster lensing analyses is
beyond the scope of this paper.

\subsection{Parametric modeling of cluster masses}\label{SS:paramodel}

In principle, we can model the cluster-galaxy weak lensing signal as a
sum of two terms, the first due to the BCG stellar component, only
important on scales below $\sim$100\hkpc, and the second due to the
dark matter halo.  Typically, the halo is modelled using the broken
power-law NFW density profile \citep{1996ApJ...462..563N}:
\begin{equation}\label{E:genprofile}
\rho(r) = \frac{\rho_s}{\left(r/r_s\right)
  \left(1+r/r_s\right)^{2}}, 
\end{equation}
where the scale radius $r_s$ is the scale at which the
  logarithmic slope, $\rmd\ln{\rho}/\rmd\ln{r}$, is equal to $-2$.
While this approach to cluster mass estimation is fairly standard,
recent work \citep{2006AJ....132.2685M, 2008MNRAS.387..536G} suggests
that the Einasto profile \citep{Einasto65},
\begin{equation}
\rho(r) =\rho_s e^{(-2/\alpha)[(r/r_s)^{\alpha}-1]}, 
\end{equation} 
(where $\alpha$ has a weak mass dependence with a value around 0.15)
may better describe the dark matter halo profiles.  We note here that
on the scales we use for modeling in this work, the two profiles agree
to within a few per cent.  Thus, the NFW profile is sufficient for our
purposes.

It is convenient to parametrise the NFW profile by two parameters,
the concentration $\cvir=\rvir/r_s$ and the virial mass \mvir.  The virial
radius $\rvir$ and $\rho_s$ can be related to $\mvir$ via consistency
relations.  The first is that the virial radius is defined such that 
the average density within it is $200\overline{\rho}$:
\begin{equation}\label{E:rvir}
\mvir = \frac{4\pi}{3}\rvir^3 \left(200\overline{\rho}\right).
\end{equation}
The second relation, used to determine $\rho_s$ from $\mvir$ and
$\cvir$, is simply that the volume integral of the density profile out
to the virial radius must equal the virial mass (though when computing
the lensing signal, we do not truncate the profiles beyond $\rvir$).
The NFW concentration is a weakly decreasing function of halo mass,
with a typical dependence as
\begin{equation}\label{E:cmrelation}
\cvir=\frac{c_0}{1+z} \left( \frac{M}{M_0} \right)^{-\beta}, 
\end{equation}
with $\beta \sim 0.1$
\citep{2001MNRAS.321..559B,2001ApJ...554..114E,2007MNRAS.381.1450N},
making this profile a one-parameter family of profiles. The
normalisation of Eq.~\eqref{E:cmrelation} depends on the nonlinear mass
(and hence cosmology), but for the typical range of models, one
expects $\cvir=5$--$8$ at $M_0=10^{14}\hmsun$.  Some more recent work
\citep{2007MNRAS.381.1450N,2009ApJ...707..354Z} suggests 
that this mass dependence levels off to
a constant concentration above some high value of mass. 
The precise value remains somewhat controversial, with $\cvir \sim 5-6$
in \cite{2007MNRAS.381.1450N,2009ApJ...707..354Z}, but some 
other analyses suggest a significantly higher value around $ \cvir \sim 7-8$ 
at $z=0$ (J. Tinker, private communication).  In addition, if one
applies the typical concentration-mass relation assumed in, e.g.,
\cite{2007MNRAS.379..317H} to very high mass clusters, one finds very small concentration values, 
e.g. $\cvir \sim 4$ at $M \sim 10^{15} \msun$.  In this
paper, we will assess the effect of assuming the wrong concentration
value in parametric mass estimates, taking $\cvir =4$--$7$
as the plausible range given the current level of uncertainties.

While we demonstrate our cluster mass estimation procedure using
stacked lensing data for which a spherical model is appropriate, one
can easily apply the same techniques using lensing data for individual
clusters.  In that case, parametric profile fitting may use a circular
average of the shear profile, or a full shear map with the inclusion
of a projected ellipticity and position angle among the fit
parameters.  Here, for simplicity, we assume the former.

There are two significant practical differences between stacked survey
data versus data for individual clusters: First, survey data are
typically available to large transverse separations, whereas data for
individual clusters are limited by the field of view (FOV) of the telescope
used for the observations.  For typical cluster redshifts in cluster
lensing analyses, $2$\hmpc\ is a typical maximum radius to which the
lensing signal can be measured.  Second, stacked lensing data
typically yields a concentration that is around the mean concentration
of the sample used for the stacking \citep{2005MNRAS.362.1451M}.  As a
result, the main uncertainty in what concentration to assume for
parametric mass estimation comes from differences between the
published concentration-mass relations from $N$-body simulations, the
uncertainty in cosmological parameters, and the uncertainty about how
baryonic cooling may have changed the halo concentration.  In
contrast, for individual cluster data the concentration is likely to
vary significantly from cluster to cluster due to the intrinsic
lognormal concentration distribution at fixed mass; this variation of
$\sim 0.15$ dex \citep{2001MNRAS.321..559B} is non-negligible compared
to the sources of systematic uncertainty about halo concentration.

In this paper, when studying the effects of parametric models on fits
for the mass, we choose to fix the halo concentration as in some
individual analyses, such as \cite{2007MNRAS.379..317H}, and some
stacked analyses, such as \cite{2008MNRAS.390.1157R}.  Other works
have fit simultaneously for a concentration and a mass \citep[e.g.,
][]{2008JCAP...08..006M,2009arXiv0903.1103O}.  In the latter case,
there is no concern about biases in the mass due to assumption of the
wrong concentration, but small biases may remain due to deviations of
the profile from NFW, and there is a loss of statistical power so that
the mass estimates become noisier.  Furthermore, if there are
systematic errors in the data (such as centroiding errors or intrinsic alignments) that do
not perfectly mimic a change in concentration, those analyses may
still find a biased result for the mass.  For the most part, we wish
to characterise systematic biases that can occur when the
concentration is fixed, but we will mention the effects of allowing it
to vary.

Finally, we note that parametric mass estimation lends itself easily
to corrections for effects such as non-weak shear and magnification
bias \citep{2006MNRAS.372..758M}.  These effects can simply be
incorporated into the model before comparing with the data.

\subsection{Non-parametric modeling of cluster masses}

Another common approach to cluster mass estimation is the
non-parametric aperture mass statistic.  In this work, we present
tests of the \zetac\ statistic \citep{1998ApJ...497L..61C}, which is
related to the $\zeta$ statistic of \cite{1994ApJ...437...56F}.
\zetac\ has been used in several recent cluster modeling papers,
including \cite{2007MNRAS.379..317H} and \cite{2009arXiv0903.1103O}.
This statistic is defined using three radii: the
first, $R_1$, is the transverse separation within which we wish to
estimate the projected mass; the second and third, $R_{o1}$ and
$R_{o2}$, define an outer annulus.  \zetac\ is equal to the mean
surface density within $R_1$ relative to that in the outer annulus:
\beq 
\zetac(R_1) =
\overline{\kappa}(R<R_1) -\overline{\kappa}(R_{o1}<R<R_{o2}), 
\eeq
where $\kappa$ is the scaled surface density or convergence,
$\kappa=\Sigma/\Sigma_c$.  The aperture mass statistic can be
measured using the observed shear $\gamma_t(R)$ using 
\begin{align}
\zetac(R_1) &= 2\int_{R_1}^{R_{o1}} \rmd\ln{R}\,\gamma_t(R) +  \label{E:zetac}\\
&\qquad\frac{2}{1-(R_{o1}/R_{o2})^2} \int_{R_{o1}}^{R_{o2}}
\rmd\ln{R}\,\gamma_t(R).\notag
\end{align}
The 2d (projected or cylindrical) mass $\mproj(R_1)$ within $R_1$ can be estimated
from \zetac\ by 
\beq 
\mproj(R_1) = \pi R_1^2 \,\Sigma_c\,\zetac(R_1). \label{E:mproj} 
\eeq

Typically $R_1$ is chosen to be either a fixed physical scale, or a
spherical over-density radius (determined either using a parametric
model to estimate the appropriate radius, or iteratively using the
aperture mass estimate from the data).  Various approaches are taken
to the second term in Eq.~\eqref{E:zetac}, which should ideally be
sub-dominant to the first given the scaling of shear with radius.  For
example, \cite{2007MNRAS.379..317H} use the parametric fits to an NFW
profile with fixed concentration parameter to estimate the amplitude
of the second term.  In contrast, \cite{2009arXiv0903.1103O} neglect
it, after choosing $R_{o1}$ to be 10--15 arcmin, depending on where in
the cluster field there appeared to be significant structures that
they wished to avoid.\footnote{M. Takada, private communication} For
their typical cluster redshifts, this choice corresponds to roughly
$2$--$2.5$ comoving \hmpc\ in transverse separation.  We will consider
the effect of both approaches in our tests below.

The aperture mass statistic is often used because of its insensitivity
to the details of the cluster mass profile.  Furthermore, because it
estimates the mass within $R_1$ using the shear on scales larger than
$R_1$, it is not very sensitive to systematics that affect the signal
in the inner parts, such as contamination by cluster member galaxies,
intrinsic alignments, and centroiding errors.  This decreased
sensitivity to systematics comes at a price, however: as shown in
Eq.~\eqref{E:zetac}, the determination of \zetac\ requires integration
over the measured shear profile in logarithmic annular bins, which can
often be quite noisy.  Our tests will help quantify the extent to
which this noisiness increases the statistical error on the mass
estimates relative to parametric modeling.

An additional disadvantage to the use of the aperture mass statistic
and the derived \mproj\ is that cosmological analyses using the mass
function, and any comparison against X-ray-derived masses, requires
the use of a 3d (enclosed) mass, \menc.  The conversion from \mproj\
to \menc\ requires the assumption of a profile, such as NFW (for which
a concentration parameter must either be assumed, or derived from
parametric fits).  This conversion factor may be derived analytically
from expressions for the enclosed \mproj\ and \menc\ as in
\cite{2000ApJ...534...34W}.  \cite{2009arXiv0903.1103O} show that the
conversion factor only weakly depends on the concentration, but for
analyses that seek to determine the mass to 10 per cent, this
dependence on concentration is still important.  A way of avoiding
this necessity would be to determine the mass function in terms of
projected masses in the simulations, rather than the typical practise
of using \menc\ within some spherical over-density; however, given that
this has not yet been done, we also test the effect of this \mproj\ to
\menc\ conversion.

\subsection{New statistic for mass estimation}\label{SS:newmassest}

As noted previously, one complication in parametric modeling of the
lensing signal \dsr\ is the sensitivity to the mass profile on small
scales, which is particularly prone to theoretical and observational 
uncertainty.  We wish to avoid sensitivity to small scales, which
comes from the first term on the right-hand side of Eq.~\eqref{E:ds},
via \avsigr\ (defined in Eq.~\ref{E:avsig}).

Thus, we must turn the lower limit of integration in Eq.~\eqref{E:avsig}
from $R=0$ to some larger scale that is not strongly affected by
small-scale systematics such as intrinsic alignments and centroiding
errors.  We refer to this new minimum scale as $R_0$, and achieve our
goal by defining the annular differential surface density (ADSD) 
\begin{align}
\ups &= \dsr - \ds(R_0)\left(\frac{R_0}{R}\right)^2 \label{E:upsdef} \\
  &= \frac{2}{R^2}\int_{R_0}^{R} \Sigma(R') R' \,\rmd R' - \sigr +
  \Sigma(R_0)\left(\frac{R_0}{R}\right)^2. \notag
\end{align}
As shown in Eq.~\eqref{E:upsdef}, by subtracting off $\ds(R_0)\left(R_0
  /R\right)^2$ from the observed lensing signal, we achieve our goal
of removing the sensitivity to scales below $R_0$.  The resulting
robustness of the analysis to systematic errors comes at the expense
of introducing slight ($\sim 10$ per cent level) anti-correlations
between the signal around $R_0$ and the signal at larger scales, plus
increased statistical errors.

For some of this paper, we model theoretical and observational
  uncertainties in \ds\ as changes in the NFW concentration parameter.
  However, as already discussed, some systematics are manifested in
  different ways (e.g., centroiding errors) that must be modeled
  rather differently.  If one truly believes that unknown
  concentration is the dominant systematic uncertainty, then the
  simplest solution would be to fit \ds\ to an NFW profile and then
  marginalize over the concentration.  Since we do not believe that
  this procedure is adequate for all theoretical and observational
  systematics, parametric modeling of \ups\ to remove all small-scale
  information is a better solution that will give more accurate mass
  estimates.  For some systematics, we will see in
  Section~\ref{S:results} that we do not, in general have to select
  $R_0$ to be completely above the affected scales, because the errors
  in $\ups$ change sign and thus nearly cancel out of the mass
  estimation, despite contributing to \dsr\ with the same sign at all
  scales.

In practise, to use the ADSD \ups\ we must estimate $\ds(R_0)$ from the data
itself.  In this work, we have tried two methods of doing so based on
fits to the following functional form for $\ds$ in the neighbourhood of $R_0$:
\begin{equation}\label{E:fittingfunc}
\ds(R) = \ds(R_0) \left(\frac{R}{R_0}\right)^{p + q(R/R_0)}
\end{equation}  
In the simpler method, $q=0$, whereas in the more complex method it is
a free parameter in the fit (which generally allows for a better fit
to broken power-law profiles such as NFW, but also increases the
statistical errors on the mass).  We primarily present results of the
latter procedure, but discuss the trade-offs between the two in
Section~\ref{S:results}.

Finally, we note that the ADSD \ups\ is well-suited not only to
estimating cluster masses, but also to cosmological studies, where the
choice of $R_0$ to be outside of the host dark matter halo virial
radius allows contributions to the lensing signal from small-scale
information to be suppressed \citep{tobiaspaper}.

\section{Simulations}
\label{S:simulations}

To obtain realistic cluster lensing profiles for our tests of mass
inference methods, we use the Z\"{u}rich horizon ``\texttt{zHORIZON}''
simulations, a suite of thirty pure dissipationless dark matter
simulations of the $\Lambda$CDM cosmology \citep{2009MNRAS.400..851S}. Each simulation models the
dark matter density field in a box of length $L=1500$\hmpc, using
$N_\text{p}=750^3$ dark matter particles with a mass of
$M_\text{dm}=5.55 \times 10^{11} \hmsun$. The cosmological parameters
for the simulations in Table \ref{tab:cosmoparam} are inspired by the
results of the WMAP cosmic microwave background experiment
\citep{2003ApJS..148..175S,2007ApJS..170..377S}. For this work, we use
eight of the thirty simulations, and probe a volume of $27
h^{-3}\mathrm{Gpc}^3$ at redshift $z=0.23$. The initial conditions
were set up at redshift $z=50$ using the \texttt{2LPT} code
\citep{1998MNRAS.299.1097S}. The evolution of the $N$ equal mass
particles under gravity was then followed using the publicly 
available $N$-body code \texttt{GADGET-II}
\citep{2005MNRAS.364.1105S}.  Finally, gravitationally-bound
structures were identified in each simulation snapshot using a
Friends-of-Friends \citep[FoF, ][]{1985ApJ...292..371D} algorithm with
linking length of $0.2$ times the mean inter-particle spacing. We
rejected halos containing fewer than twenty particles, and identified
the potential minimum of the particle distribution associated with the
halo as the halo centre. We note that using the FoF halo finder might
cause some problems with the halo profile, since FoF tends to link
together nearby halos.  In total, we identify halos in the mass
range $1.1 \times 10^{13}\hmsun\leq \mvir \leq 4 \times 10^{15}
\hmsun$.

\begin{table}
	\centering
	\begin{tabular}{llllll}
	\hline
	\hline
	$\Omega_\text{m}$&$\Omega_\Lambda$&$h$&$\sigma_8$&$n_s$&$w$\\
	\hline
	$0.25$ &$0.75$&$0.7$&$0.8$&$1.0$&$-1$\\
	\hline	 
	\hline
	\end{tabular}
	\caption{Cosmological parameters adopted for the simulations: matter
density relative to the critical density, dark energy density
parameter, dimensionless Hubble parameter, matter power spectrum
normalisation, primordial power spectrum slope, and dark energy equation of state
$p=w \rho$.}
	\label{tab:cosmoparam}
\end{table}

\subsection{Calculation of the signal}

We calculate the spherically-averaged correlation function in the
simulations using direct counts of mass particles in spherical shells
about the halo centres of the cluster stack, $N_\text{cl,m}(r_i)$. Our
estimator for the correlation function is
\begin{equation}
 \xi_\text{cl,m}(r_i)=\frac{N_\text{cl,m}(r_i)}{N_\text{cl,m}^\mathrm{(rand)}(r_i)}-1,
\end{equation} 
where $N_\text{cl,m}^\mathrm{(rand)}(r_i)=N_\text{cl} N_\text{m}
V_\text{shell}/V_\text{box}$ is the expected number of pairs for a
purely random sample (for $N_\text{cl}$ and $N_\text{m}$ defined as
the total number of clusters and matter particles in the box, respectively), and $V_\text{shell}=4\pi(r_{i+1}^3-r_i^3)/3$ is
the volume of the spherical shell at $r_i$. To reduce the
computational cost of this calculation, we dilute the dark matter
density field by a factor of 24, using only $20\times 10^{6}$ dark
matter particles. We have confirmed the convergence of this procedure.

\subsection{Correction for resolution effects}

Despite the large dynamical range of our simulations, our
resolution is still limited on small scales. The force softening
length was set to $70$\hkpc, so our results may not be reliable for
$r\lesssim 200$\hkpc. This resolution problem limits our ability to
predict the excess surface mass density \dsr\ on small scales, since
this quantity is affected by the average over the correlation function
on even smaller scales.  Therefore, to correct for this problem, we
continue the profile toward small scales using the NFW profile as
follows:
\begin{equation}
1+\xi_\text{cl,m}^\mathrm{(stitch)}(r)=\begin{cases}
  \rho_\text{cl,m}^\mathrm{(NFW)}(r)/\rhob,  & \text{for } r<r_\text{stitch}\\
  \rho_\text{cl,m}^\mathrm{(sim)}(r)/\rhob, & \text{for } r\geq r_\text{stitch}
\end{cases}
\end{equation}
We used the combinations $(r_\text{stitch}=0.2$\hmpc, $\cvir=5)$ and
$(r_\text{stitch}=1.0$\hmpc, $\cvir=7)$.
\par
Virial radii and masses are calculated by imposing the constraint
\begin{equation}
 \frac{3}{\rvir^3 \delta}\int_0^{r_\text{vir}} (r')^2\,\rmd r' \
\bigl[1+\xi_\text{cl,m}(r')\bigr]=\frac{3 \mvir}{4 \pi \rvir^3 \bar
\rho \delta} =1.
\end{equation}
The over-density of halos is assumed to be $\delta=200$ times the
background density.  The profile is then spline fitted and integrated
along the line of sight, over separations $-50\le\chi\le
50$\hmpc\ from the
cluster.   

%
%
%
%

\section{Data}
\label{S:data}

The SDSS \citep{2000AJ....120.1579Y} imaged roughly $\pi$ steradians
of the sky, and followed up approximately one million of the detected
objects spectroscopically \citep{2001AJ....122.2267E,
  2002AJ....123.2945R,2002AJ....124.1810S}. The imaging was carried
out by drift-scanning the sky in photometric conditions
\citep{2001AJ....122.2129H, 2004AN....325..583I}, in five bands
($ugriz$) \citep{1996AJ....111.1748F, 2002AJ....123.2121S} using a
specially-designed wide-field camera
\citep{1998AJ....116.3040G}. These imaging data were used to create
the cluster and source catalogues that we use in this paper.  All of
the data were processed by completely automated pipelines that detect
and measure photometric properties of objects, and astrometrically
calibrate the data \citep{2001adass..10..269L,
  2003AJ....125.1559P,2006AN....327..821T}. The SDSS was completed
with its seventh data release \citep{2002AJ....123..485S,
  2003AJ....126.2081A, 2004AJ....128..502A, 2005AJ....129.1755A,
  2004AJ....128.2577F,
  2006ApJS..162...38A,2007ApJS..172..634A,2008ApJS..175..297A,2009ApJS..182..543A}.

In this paper, the only data that we use are the maxBCG cluster
lensing data previously analysed in \cite{2008JCAP...08..006M}.
Because the data were described there in detail, here we simply give a
brief summary.

The parent sample from which our lens samples were derived consists of
13~823 MaxBCG clusters
\citep{2007ApJ...660..221K,2007ApJ...660..239K}, identified by the
concentration of galaxies in colour-position space using the well known
red galaxy colour-redshift relation \citep{2000AJ....120.2148G}. The
sample is based on 7500 square degrees of imaging data in SDSS.  There
is a tight mass-richness relation that has been established using
dynamical information \citep{2007ApJ...669..905B} and weak lensing
\citep{2007arXiv0709.1159J,2008JCAP...08..006M,2008MNRAS.390.1157R}
across a broad range of halo mass.  The redshift range of the maxBCG
sample is $0.1<z<0.3$; within these redshift limits, the sample is
approximately volume-limited with a number density of $3\times 10^{-5}
(h/\mbox{Mpc})^3$, except for a tendency towards higher number density
at the lower end of this redshift range \citep{2008MNRAS.390.1157R}.
In this paper, we use scaled richness in red galaxies above $0.4L_*$
within $R_{200}$, known as \nt, as a primary tracer of halo mass.  For
the data in \cite{2008JCAP...08..006M} that we use here, the richness
range is $12\le\nt\le 79$ divided into six bins ($12\le\nt\le 13$,
$14\le\nt\le 19$, $20\le\nt\le 28$, $29\le\nt\le 39$, $40\le\nt\le
54$, and $55\le\nt\le 79$).

The source sample with estimates of galaxy shapes is the same as that
originally described in \citet{2005MNRAS.361.1287M}.  This source
sample has over 30 million galaxies from the SDSS imaging data with
$r$-band model magnitude brighter than 21.8, with shape measurements
obtained using the REGLENS pipeline, including PSF correction done via
re-Gaussianization \citep{2003MNRAS.343..459H} and with cuts designed
to avoid various shear calibration biases.  The overall calibration
uncertainty due to all systematics was originally estimated to be
eight per cent \citep{2005MNRAS.361.1287M}, though the redshift
calibration component of this systematic error budget has recently
been decreased due to the availability of more spectroscopic data
\citep{2008MNRAS.386..781M}.  The absolute mass calibration is not a
critical issue for this paper, in which we study the changes in
estimated mass for a given observed signal when using different
estimation procedures.

\section{Results}\label{S:results}

\subsection{Purely analytical profiles}\label{SS:theorresults}

In this subsection, we add realistic levels of noise to pure NFW
profiles to create simplified mock cluster density profiles. The
profiles that we use have $\log_{10}[{h\mvir/\msun}] = 14.0$ and $14.8$,
with $\cvir=4$ and $\cvir=7$ (see properties of these profiles listed in
Table~\ref{T:testprofiles}).  Using these profiles, we can test the
dependence of parametric and non-parametric modeling on assumptions
about the NFW concentration parameter.  We caution that these profiles
cannot be used to test for the effects of deviations from an NFW
profile on the extracted masses when fitting assuming NFW profiles, or
for the effects of large-scale structure contributions to the lensing
signal. These are discussed in the next subsection. 
\begin{table*}
\begin{center}
\caption{Properties of cluster lensing profiles, both analytical (pure
  NFW) and those from $N$-body simulations.  We show the mean number
  density of the sample for the mass-selected samples from $N$-body
  simulations; the virial mass and radius \mvir\ and \rvir (exact
  value for the pure NFW profiles, and the ensemble mean for the
  samples from $N$-body simulations); the
  analytical profiles used for resolution corrections of the $N$-body
  simulations; and the best-fitting NFW profiles when fitting the
  simulation lensing signals \dsr\ for scales $0.2\le R \le 2$\hmpc.  \label{T:testprofiles}}
\begin{tabular}{cccccc}
\hline\hline
 & & & & \multicolumn{2}{c}{Best-fit NFW parameters} \\
$\overline{n}$ & $\mvir$ & $\rvir$ & stitching & $\mvir$ & $\cvir$ \\
$\!\!\![10^{-6} (h/\text{Mpc})^3]\!\!\!$ & $\!\!\![10^{14} \hmsun]\!\!\!$ & $\!\!\![$\hmpc$]\!\!\!$ & &
$\!\!\![10^{14}\hmsun]\!\!\!$ & \\
\hline
\multicolumn{6}{c}{Pure NFW profiles, $\cvir=4$ and $7$} \\
- & $1.0$ & $1.2$ & - & - & - \\
- & $6.3$ & $2.2$ & - & - & - \\
\hline
\multicolumn{6}{c}{$N$-body simulation profiles} \\
$0.25$ & $7.9$ & $2.4$ & $\!\!\!\cvir=7$ at $1$\hmpc$\!\!\!$ & $7.9$ & $6.6$ \\
$0.25$ & $7.5$ & $2.3$ & $\!\!\!\cvir=5$ at $0.2$\hmpc$\!\!\!$ & $7.8$ & $4.6$ \\
$2$ & $4.2$ & $1.9$ & $\!\!\!\cvir=7$ at $1$\hmpc$\!\!\!$ & $4.2$ & $6.6$ \\
$2$ & $4.0$ & $1.9$ & $\!\!\!\cvir=5$ at $0.2$\hmpc$\!\!\!$ & $4.3$ & $4.6$ \\
$16$ & $1.6$ & $1.4$ & $\!\!\!\cvir=7$ at $1$\hmpc$\!\!\!$ & $1.7$ & $6.5$ \\
$16$ & $1.6$ & $1.4$ & $\!\!\!\cvir=5$ at $0.2$\hmpc$\!\!\!$ & $1.7$ & $4.5$ \\
\hline
\hline
\end{tabular}
\end{center}
\end{table*}

These values of concentration were selected as the extremes of the
variation allowed with cosmology, and with the various determinations
of the concentration-mass relation in the literature, including recent
results suggesting that the concentration stops decreasing with mass
at the high-mass end \citep{2007MNRAS.381.1450N,2009ApJ...707..354Z}.  In addition, we
consider that baryonic effects may increase the concentration of the
dark matter profile (for an extreme example, see
\citealt{2008ApJ...672...19R}).  Furthermore, for individual cluster
lensing analyses, we must consider the fact that dark matter halos
exhibit a large scatter in concentration (0.15 dex,
\citealt{2001MNRAS.321..559B}), so the variation we have used is not
as extreme in this case, as it may be for a stacked cluster analysis.
The change in concentration from $4$ to $7$ is less than $2\sigma$ of
this intrinsic scatter.

To generate the profiles, we begin with the cluster halo
  density profile $\rho_\text{cl}(r)$, which is defined in very narrow
  logarithmic (3d) radial bins.  We then numerically integrate this
  profile along the line-of-sight, for comoving line-of-sight
  separations $|\chi|\le\rvir$, to define $\Sigma(R)$ in very narrow
  logarithmic bins in transverse separation $R$.  We calculate
  \avsigr\ by converting the integral in Eq.~\eqref{E:avsig} to a
  summation.  $\ds(R)$ can then be computed directly from
  $\avsigr-\Sigma$. 

To make this theoretical signal, defined in very narrow bins without
  any noise, look like an observed signal, we then do the following.
  First, we use a spline to determine the values of \ds\ at the center
  of the bins in $R$ used to calculate the real signal for maxBCG
  clusters in \cite{2008JCAP...08..006M}.  Second, we choose a cluster
  richness subsample from that paper with roughly comparable mass to
  the theoretical signal we are using.  We estimate a power-law function
  for the (bootstrap-determined) errors as a function of radius from
  our selected cluster subsample, to avoid the influence of any 
  noise in the determination of the covariances.  We use this
  power-law to assign a variance to the theoretical signal as a
  function of transverse separation.  Finally, since the signal in the
  different radial bins were found to be nearly uncorrelated for all scales
  used in that paper, we add noise to our theoretical signals using a
  Gaussian distribution with a diagonal covariance matrix.  This
  procedure was performed 1000 times to generate 1000 realizations of
  the lensing data.  For context, the input level of noise is typically
  sufficient to achieve $\sim 20$ per cent statistical uncertainty on
  the best-fit masses at the $1\sigma$ level, when using \ds\ with
  $R<4$\hmpc\ to fit for the mass.

\begin{figure*}
\begin{center}
\includegraphics[width=5in,angle=0]{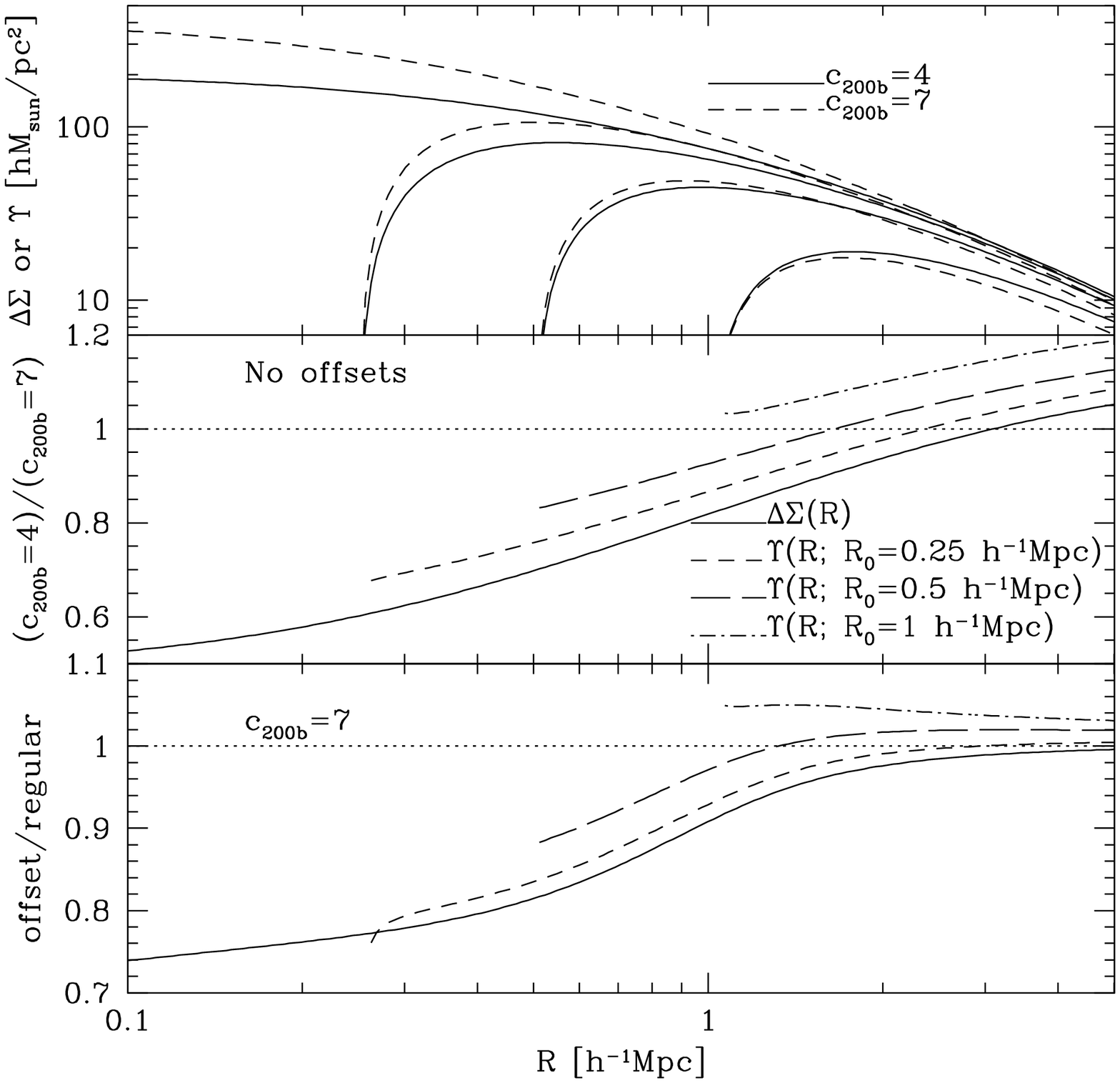}
\caption{\label{F:theorsig}Top panel: from top to bottom, we show
  \dsr\ and \ups\ with $R_0=0.25$, $0.5$, and $1$\hmpc.  The solid
  lines are for $\cvir=4$ and the dashed lines are for $\cvir=7$; in
  both cases, $\log_{10}[{h\mvir/\msun}]=14.8$.  Middle panel: without
  inclusion of centroid offsets, we show the ratio
  of these four quantities for $\cvir=4$ versus $\cvir=7$, where the 
  line types indicate which quantity is used to construct the ratio,
  and the horizontal dotted line indicates a ratio of 1.  Bottom
  panel: assuming $\cvir=7$, we show the ratio of these four
  quantities when including centroiding offsets versus not, with the
  same line styles as in the middle panel.}
\end{center}
\end{figure*}

The input lensing signals \dsr\ and \ups\ (before the addition of noise) with several $R_0$ values
are shown in Fig.~\ref{F:theorsig} for the higher mass value,
$\log_{10}[{h\mvir/\msun}]=14.8$.  Since we will also test the effect of
centroiding errors, which were discussed in detail in
Section~\ref{SS:challenges-theory}, we apply the offset model from
\cite{2007arXiv0709.1159J}.  For offset fractions, we have chosen 20
per cent for this mass scale; for $\log_{10}[{h\mvir/\msun}]=14.0$, we will
use 35 per cent (roughly in accordance with the trends with richness
in that paper).

As expected, \dsr\ for $\cvir=7$ is higher than that for $\cvir=4$ on
small scales; the radius at which they cross over is relatively large
because \dsr\ includes information from $\Sigma(R)$ for small $R$.  For
the 3d $\rho(r)$, the cross-over radius is within the virial radius by
necessity, since the masses are the same.  As we increase $R_0$ in
\ups, the trend going from $\cvir=4$ to $\cvir=7$ gets less
pronounced, because even though \dsr\ is larger on small scales for
$\cvir=7$, that also means the value that is subtracted off to obtain
\ups\ is larger.  Thus, by the time we reach $R_0=1$\hmpc, \ups\ is
actually higher for $\cvir=4$ than for $\cvir=7$ for all $R>R_0$.  

As shown in the bottom panel, the effect of centroiding errors is
quite pronounced on \dsr.  The characteristic scale of the offsets is
$0.42$\hmpc, and the signal is noticeably suppressed out to three times
this scale.  The use of \ups\ ameliorates this effect, and it even gets reversed for larger $R_0$, similar to what
happens with the different concentration values.  While for \dsr, the
offsets cause suppression of the signal for all affected scales, for
\ups, the signal is suppressed on smaller scales and elevated on
larger scales, which suggests that biases in parametric mass modeling
due to these offsets may be smaller because the small and large scale
changes in sign may cancel out.


\subsubsection{Parametric modeling}\label{SSS:analytic-para}

In this section, we begin by fitting the pure NFW lensing signals for
$\log_{10}[{h\mvir/\msun}]=14.8$ to pure NFW profiles.  This procedure
allows us to assess the systematic uncertainty due to the assumption
of a fixed concentration when using various parametric fit procedures.
For each noise realisation, we attempted to determine a mass using
several fitting procedures:

\begin{itemize}
\item Assuming an NFW profile with $\cvir=4$ and $\cvir=7$.
\item Using \dsr\ with minimum fit radii (\rmin) values ranging from $0.1$
  to $2$ \hmpc, maximum fit radii of $\rmax=1$, $2$, and $4$ \hmpc.
\item Using \ups\ with $R_0=0.25$, $0.5$, and $1$ \hmpc, again with a
  variety of \rmin\ values (always with $\rmin > R_0$).  The value of
  $\ds(R_0)$ was determined on each noisy realisation rather than from
  the well-determined mean over those scenarios, consistent with a
  real measurement for which we only have one observation of the
  lensing signal for a given sample.  The estimation was done by
  fitting the data to the three-parameter functional form in
  Eq.~\eqref{E:fittingfunc} from $0.1<R<0.5$, $0.3<R<1$, and
  $0.7<R<1.3$\hmpc\ for $R_0=0.25$, $0.5$, and $1$ \hmpc,
  respectively.
\end{itemize}

In detail, the fits to \dsr\ are performed via $\chi^2$ minimization
in comparison with theoretical signals that were generated via the
procedure described at the start of Section~\ref{SS:theorresults}.
Thus, for each of the lensing signal realizations $j$, denoted
$\ds_{j}^{\text{(data)}}(R_i)$ (for bins in transverse separation with index
$i$ such that $\rmin\le R_i\le \rmax$) with noise variance
$\sigma^2(\ds_{j}(R_i))$, we use the Levenberg-Marquardt algorithm
\citep{levenberg,marquardt,1992nrca.book.....P}
to find the NFW profile mass that minimizes
\beq\label{E:chi2}
\chi^2_j = \sum_i \frac{[\ds_{j}^{\text{(data)}}(R_i)-\ds^\text{(model)}(R_i|\mvir,\cvir)]^2}{\sigma^2(\ds_{j}(R_i))}
\eeq
at fixed \cvir.

The fits to \ups\ require an additional step: the conversion
  of both the theoretical signals ($\ds^\text{(model)}$, defined without
  noise in very narrow bins in $R$) and the mock data
  ($\ds^\text{(data)}$, defined in realistically broad bins with added
  noise) from \dsr\ to \ups.  In practice, the theoretical signal is
  defined such that we can very accurately interpolate to 
  determine the value of $\ds(R_0)$, which is then used to construct
  \ups\ directly using Eq.~\eqref{E:upsdef}.  For the noisy mock data, we
  must use a different procedure.  We fit to \dsr\ to
  estimate $\ds(R_0)$ using Eq.~\eqref{E:fittingfunc}, so that \ups\
  can be constructed.  We will shortly discuss more details of this
  procedure, because we find that the exact way of getting $\ds(R_0)$ is important: some
  methods introduce a bias on the mass, others add extra
  noise, neither of which is desirable.  Once \ups\ is determined for
  the mock signals, we then determine its  covariance matrix using the
  distribution of values for all datasets.  Finally, we minimize the $\chi^2$ function for
  each mock realization using Eq.~\eqref{E:chi2} with \ups\ in place
  of \dsr.

\begin{figure*}
\begin{center}
\includegraphics[width=5.5in,angle=0]{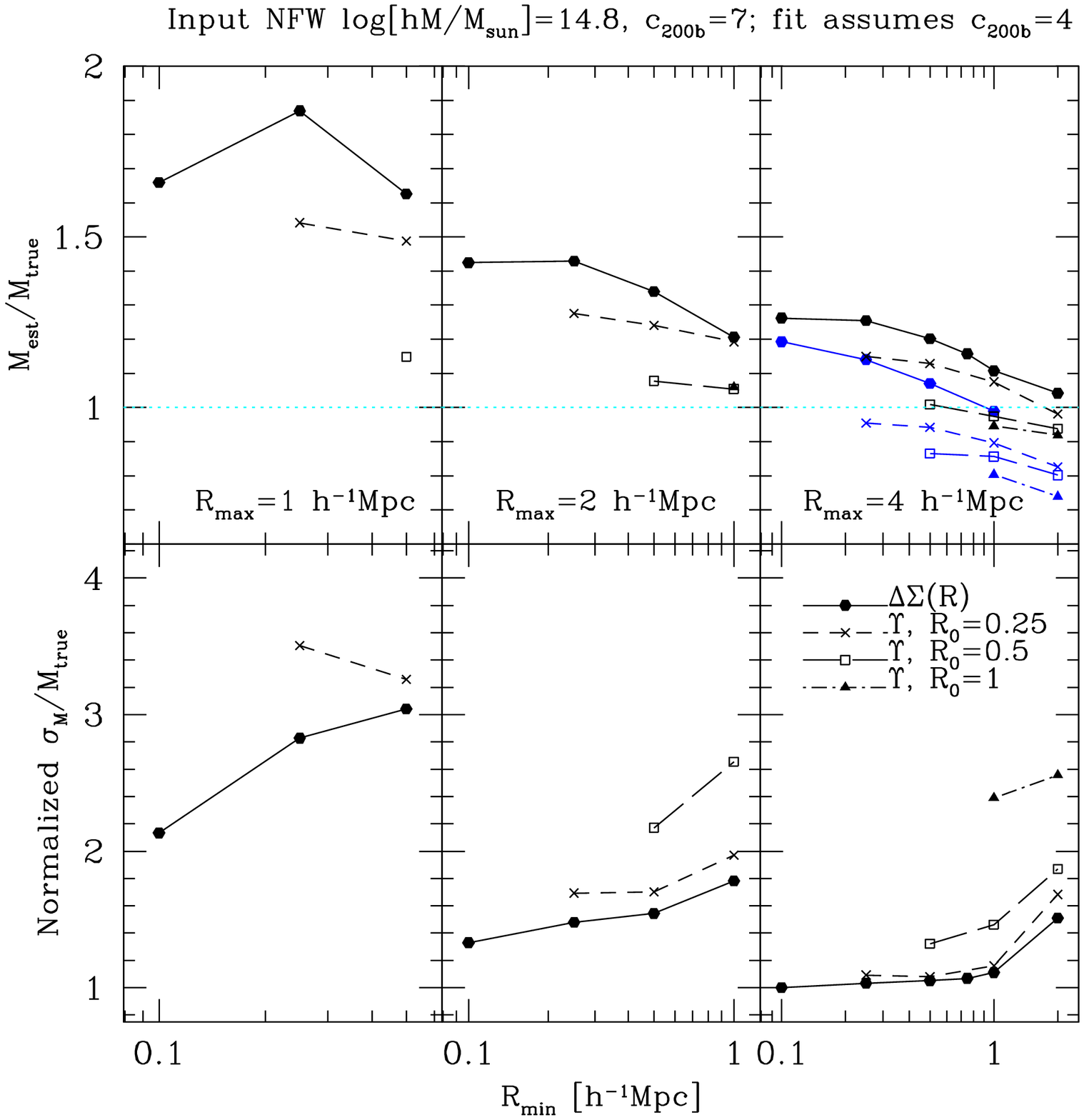}
\caption{\label{F:theorresults}Results of parametric mass fits on
  noisy realisations of pure NFW profiles, with input
  $\log_{10}[{h\mvir/\msun}] = 14.8$ and $\cvir=7$, but $\cvir=4$ assumed
  in the fits.  The top and bottom rows show the ratio $\mest/\mtrue$
  and the statistical error $\sigma(\mest)/\mtrue$, respectively.  The
  latter is shown normalised to the minimum value of
  $\sigma(\mest)/\mtrue\sim 0.2$, which is obtained for the fit using the
  maximum information, $\dsr$ with $\rmin=0.1$ and $\rmax=4$\hmpc.  The
  results are shown for various fitting methods (indicated with
  various line and point types shown on the plot) as a function of the
  minimum fit radius \rmin.  From left to right, the panels show
  increasing $\rmax$ values of 1, 2, and 4\hmpc.  On the upper rightmost
  panel, the thin (blue) lines and points show the corresponding results for
  the $\log_{10}[{h\mvir/\msun}] = 14.0$ profile.}
\end{center}
\end{figure*}

We then examined the distribution of best-fit masses for the $1000$
noise realisations to find the mass at the 16th, 50th (median), and 84th
percentile.  We define the spread in the masses, $\sigma_M$, as being half the
difference between the 84th and 16th percentile (which would be the
standard deviation for a Gaussian distribution). The mass
distributions are sufficiently close to Gaussian that using the mean
rather than the median, and using the standard deviation directly,
would not change the plots substantially.  The median best-fitting
mass \mest\ relative to the input mass \mtrue, and the spread in the
best-fitting masses, are shown for both input profiles and each fit
method as a function of \rmin\ in Fig.~\ref{F:theorresults}.  The
criterion that we apply when selecting a robust mass estimator is that
the ratio $\mest/\mtrue$ should not depend strongly on the input or
output \cvir (though a systematic offset independent of input and output
\cvir\ is acceptable, since simulations can be used to correct for
it).

We begin by considering the trends in the ratio $\mest/\mtrue$ with
fitting method.  When assuming $\cvir=4$ while fitting to the profile
with true $\cvir=7$, as shown in Fig.~\ref{F:theorresults}, the fits
to \ds\ in the upper right panel with $\rmax=4$\hmpc\  
 give $\sim 25$ per cent overestimation of
the mass for $\rmin \le 0.5$\hmpc, improving to 3 per cent with
$\rmin=2$\hmpc\ (with, however, a doubling of the statistical error).
The mass is overestimated in this case because for the majority of
the radial range used for the fitting, the lensing signal for
$\cvir=4$ for this mass is below that for $\cvir=7$
(Fig.~\ref{F:theorsig}), so the fitting routine compensates for the
discrepancy by returning a higher mass.  This trend of overestimated
masses is decreased and eventually even reversed in sign for \ups\ as
we increase $R_0$, for reasons that are clear from
Fig~\ref{F:theorsig}.   The reverse situation, with input $\cvir=4$ and assumed $\cvir=7$,
leads to biases $\mest/\mtrue$ that are the inverse of the biases
shown in Fig.~\ref{F:theorresults}, so we do not show this case in the
figures.  As shown, when using \ups\ with $\rmin=R_0$, the
statistical error increases over the minimum possible value from the
\dsr\ fits by factors
of 1.14, 1.32, and 2.25 when using $R_0=0.25$, $0.5$, and $1$\hmpc,
respectively.  

When fitting \ups\ for
all $R_0$ and \rmin, if we use a power law to fit for
$\ds(R_0)$ (i.e., $q=0$ in Eq.~\eqref{E:fittingfunc}), then \mest\ is
consistently $\sim 3$--$5$ per cent above \mtrue even if the correct
concentration is assumed in the fit.
This overestimation of the mass occurs because the data are not
consistent with a power-law.  Due to the trend of the signal with
radius, the power-law fit tends to underestimate $\ds(R_0)$, thus
overestimating \ups\ and therefore \mest.  However, we find that a
full three-parameter fit significantly increases the noise, so we
instead use a two-step procedure: we first fit with fixed $q=0$ in
Eq.~\eqref{E:fittingfunc} to get a mass, then we use the best-fitting
signal to estimate $q$ at $R_0$, and use that fixed $q$ value for a
second two-parameter fit for $\ds(R_0)$ which is used for a second fit
to \ups\ to get the mass. For the remainder of this work, we present results using
that fitting procedure in order to best estimate the mass without
increasing the noise too much. 

Our criterion for a robust mass estimator on stacked cluster lensing
data is that it should have systematic error that is relatively
independent of the input $\cvir$ or the assumed $\cvir$ for the fit,
at least when compared to the size of the statistical error.
However, this robustness should not be achieved at the expense of too
large an increase in the statistical error.  As shown, the fits to
\dsr\ do not satisfy our robustness criterion, because assuming the
wrong concentration can lead to a systematic error that is tens of per
cent for reasonable \rmin.  
\ups\ with $R_0=0.25$\hmpc\ improves somewhat on \dsr\ in this
regard, and for $\rmin=1$\hmpc\ achieves a good combination of low
systematic error and only a small increase in statistical error.
\ups\ with $R_0=0.5$\hmpc\ satisfies our criterion for robustness when
using $\rmin=R_0$ while increasing the error by about 20 per cent.  
A value of $R_0=1$\hmpc\ erases too much
information and doubles the statistical errors.  For
individual cluster lensing, the criterion for a robust mass estimator
may differ, since if one adds many more clusters then the statistical
error may further decrease below the systematic error, so an even
smaller systematic error is required.

The systematic errors shown here may be overly pessimistic for stacked
data, given the wide variation in concentration that was allowed
relative to what is seen in $N$-body simulations. However, several
other systematics discussed in Sections~\ref{SS:challenges-theory}
and~\ref{SS:challenges-obs} can mimic a change in concentration, such
as baryonic effects.  Thus, it is only reasonable that we should
consider a broader range of concentrations than in the $N$-body
simulations.  When considering a narrower range, such as
$4<\cvir<5$, the biases in the masses when fitting to \ds\ with a
  fixed concentration are
  typically of order 10 per cent, or $\lesssim 2$ per cent when fitting to \ups. For individual cluster
lensing data, given the large lognormal scatter in concentration seen
in simulations, these systematic errors we quote are not overly
pessimistic.  Furthermore, at this level of signal to noise, the fit
$\chi^2$ values are perfectly acceptable even for the wrong value of
concentration, so goodness-of-fit cannot be used to tell whether there
is a systematic error.

In the upper right panel of Fig.~\ref{F:theorresults}, there are thin (blue) lines
corresponding to a lower mass model that can be used to assess the
mass-dependence of these systematic biases. As shown, the mass
overestimation when fitting to \dsr\ is not as severe for the lower mass cluster as for the
higher mass cluster at fixed \rmin (because the strongly
concentration-dependent part of the inner profile has moved to smaller
radii). The virial radius for this mass is about 1.85 smaller than for the 
higher mass model, suggesting that the choice of $R_0$ should be mass dependent, with the 
optimal value of 15--25 per cent of the virial radius.  
In practise,
this relation between the virial radius and $R_0$ could be achieved iteratively by choosing some default value of $R_0$,
fitting with that value of $R_0$, and then using the resulting
best-fit mass to choose a more appropriate value of $R_0$ via
\begin{equation}
R_0 = (0.25 \text{\hmpc}) \left(\frac{\mvir}{10^{14}\hmsun}\right)^{1/3}.
\end{equation}
Here we have assumed $\Omega_m=0.25$ and a spherical overdensity of
$200\rhob$, and use comoving coordinations.   

We also note that the fitted masses are weakly cosmology-dependent.
For a fixed density profile, the mass that we estimate depends on the
assumed \omegam, with $\mvir\propto \omegam^{-0.25}$ (we confirmed this scaling for the limited range of $0.2\le\omegam\le 0.3$).  The 
\omegam\ dependence has two sources: first, we rescale the transverse
separation and signal amplitude to account for the \omegam\ dependence
of the distance measures used to convert $\theta$ and $\gamma_t$ to
$R$ and \ds, and second (and more significantly), the halo mass
definition changes since we use a spherical over-density of
$200\overline{\rho}$.  Thus, for higher \omegam, the over-density we
use is larger, which reduces the mass and virial radius, also
decreasing the concentration \cvir\ since the scale radius is held
fixed.

While stacked cluster lensing analyses from large surveys can provide
cluster lensing data to tens of \hmpc, individual cluster lensing
analyses that are not survey-based typically have a limit of
$\rmax=1$--2\hmpc\ depending on the cluster redshift and telescope
field of view.  Consequently, we also explore the dependence of our
results on the maximum scale used for the fits.  Based on
Fig.~\ref{F:theorsig}, we expect that the biases will be even higher
in this case, since when restricting to smaller scales the differences
between the lensing profiles \dsr\ are more pronounced for the
different values of \cvir.

The results of this test are shown only for the $\log_{10}[{h\mvir/\msun}]
= 14.8$ and $\cvir=7$ profile, with assumed $\cvir=4$, in the
different columns of Fig.~\ref{F:theorresults}.  
As expected, when we decrease \rmax\ (moving right to
left across the figure), the systematic errors increase fairly
drastically.  For $\rmax=1$\hmpc, the best we can achieve for the
fitting methods tested here is with \ups\ with $R_0=0.5$\hmpc, and
even that method has a 25 per cent systematic error.  For
$\rmax=2$\hmpc, \ups\ with $R_0=0.5$\hmpc\ gives several per cent systematic
errors for both $\rmin=0.5$ and $1$\hmpc.  It is clear that the existence
of data to $\rmax=4$\hmpc\ ($\approx 2r_{200b}$) is very helpful in
decreasing the systematic and statistical errors.

These results suggest that the choice of mass estimator may depend on
the maximum scale to which the lensing data can be measured for a
given dataset.  If $1$\hmpc\ is the maximum scale for which data is
available, then truly robust parametric measures of mass may be
difficult to find; in the next section, we explore whether
non-parametric measures may be better than parametric ones in this
case.  For larger values of \rmax, \ups\ with $R_0=0.5$\hmpc\ seems
adequate from the perspective of minimising the combination of
systematic and statistical error.  

We next consider the effect of cluster centroiding errors, which were
discussed in detail in Section~\ref{SS:challenges-theory}.  Note that
our results here are more general than that particular systematic error, since
several observational systematics in
Section~\ref{SS:challenges-obs} have a similar form. We use the
signals with $\cvir=4$ and $7$ for both $\log_{10}[{h\mvir/\msun}]=14.0$
and $14.8$, and apply the offset model from \cite{2007arXiv0709.1159J}
as described in the beginning of Section~\ref{SS:theorresults}.  It is
important to note that this is only one example of how photometric
errors in imaging data can cause centroiding errors for the cluster
catalogue.

\begin{figure*}
\begin{center}
\includegraphics[width=5.25in,angle=0]{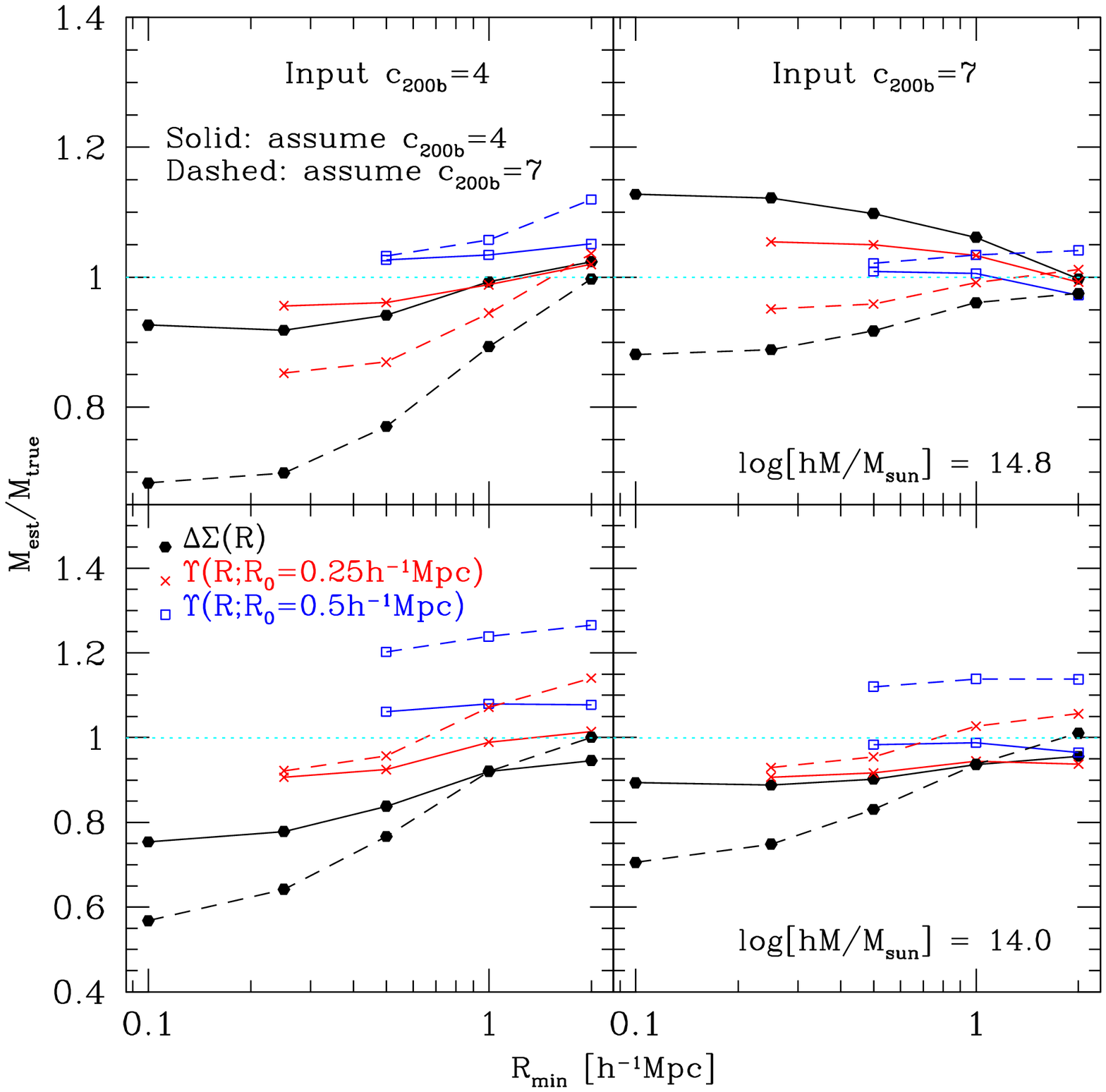}
\caption{\label{F:showoffset}Here, we show the ratio $\mest/\mtrue$
  for the pure $\cvir=4$ (left column) and $\cvir=7$ (right column)
  NFW models with $\log_{10}[{h\mvir/\msun}]=14.0$ (bottom row) and $14.8$
  (top row) after including effects of centroiding errors in the mock
  data.  Results for the various fitting methods are shown as a
  function of the minimum fit radius \rmin, for fixed $\rmax=4$\hmpc.
  The different point styles and colours as indicated on the plot show
  what type of fitting was done (\dsr\ or \ups\ with various $R_0$
  values); the different line types (solid versus dashed) indicate
  which value of \cvir\ was assumed.  The dotted horizontal lines
  indicate a ratio of $1$, the ideal unbiased case.}
\end{center}
\end{figure*}

In Fig.~\ref{F:showoffset}, we show the results of the NFW mass fits to
the profiles, with this offset distribution imposed on the data but
ignored in the fit. 
Because Fig.~\ref{F:theorresults} suggested that using \ups\ with
$R_0=1$\hmpc\ degrades the $S/N$ unacceptably, we have only shown
results for fits to \dsr\ and for \ups\ with $R_0=0.25$ and
$0.5$\hmpc.  As shown, for the higher mass model, for the input
$\cvir=4$ models, even when the correct \cvir\ is assumed in the fit
to \ds, the best-fitting masses are reduced by 5--25 per cent (lower
mass) and by up to 7 per cent (higher mass) depending on \rmin.  For
the higher mass model, we find that \ups\ with $R_0=\rmin=0.5$\hmpc\
gives fairly consistent results regardless of the input and assumed
concentration.  For the lower mass model, \ups\ with $R_0=\rmin=0.25$
\hmpc\ gives the most consistent results regardless of assumed \rmin.
Moving to the right column of this figure, for input $\cvir=7$, we see
that even with the correct assumed \cvir, fitting with \dsr\ can lead
to underestimated masses by up to 30 per cent (lower mass) or 10 per
cent (higher mass) depending on \rmin.  As for the input $\cvir=4$
model, we find that the fitting technique and minimum scale that is
most independent of assumed \cvir\ is \ups\ with $R_0=\rmin=0.5$\hmpc\
and $0.25$\hmpc\ for higher and lower mass scales, respectively.  The
ability of \ups\ to robustly estimate masses even with these
centroiding errors is a consequence of what we have noted in the
bottom panel of Fig.~\ref{F:theorsig}, that the centroiding errors
lead to biases in \ups\ that change sign at some intermediate scale,
so their effects approximately cancel out.

One important point raised by Fig.~\ref{F:showoffset} is that the mass
estimates using $\cvir=4$ (assumed) are less affected by centroid offsets.
This finding results from the fact that with a low concentration, the model
already includes a relatively low level of mass in the inner cluster
regions, and therefore is less affected than a higher concentration
halo.  Thus, it may be advantageous to assume a concentration at the 
low end of the expected range when fittings to \ups\ in scenarios
involving possibly substantial offsets of the chosen BCG from the true cluster center.


\subsubsection{Non-parametric modeling}\label{SSS:resnonpara}

In this section, we use the same noisy realisations of theoretical
cluster profiles as in the previous section, but 
we estimate masses using the aperture mass statistic \zetac.  In this case,
we begin with the NFW profile with $\log_{10}[{h\mvir/\msun}] = 14.8$ and
$\cvir=7$.  We try various options for the different aspects of this
analysis:

\begin{itemize}
\item Varying $R_1$ (the radius below which we are trying to estimate
  the enclosed mass, using the shear above that radius) between three
  values: $0.275$, $0.5$, and $1.1$ \hmpc
\item Varying $R_{o1}$ between three values: $1.1$ and $2.0$\hmpc.
\item Varying $R_{o2}$ between two values: $2$ and $4$\hmpc\ 
  (maintaining at all times the strict hierarchy $R_1 <
  R_{o1} < R_{o2}$).
\item Neglecting the second term in Eq.~\eqref{E:zetac} as in
  \cite{2009arXiv0903.1103O}, and estimating it using the best-fit NFW
  profile with some assumed concentration, as in
  \cite{2007MNRAS.379..317H}.   We do not test the case in which
  the integral from $R_{o1}$ to $R_{o2}$ may be done analytically,
  because often for individual cluster lensing studies, this is not
  even possible since $R_{o2}$ is outside the field of view.  With
  survey data or mosaic telescope data, the signal may indeed be
  measured to $R_{o2}$, but it is typically quite noisy on those large
  scales, so this procedure would introduce even more noise into the
  estimated masses.
\item Assuming $\cvir=4$ and $7$ whenever a profile assumption is
  necessary: for the estimate of the second term in
  Eq.~\eqref{E:zetac}, and for the conversion from $\mproj(<R_1)$ to
  the 3d \mvir.
\end{itemize}

The procedure is as follows.  We use the (noisy) realizations of the
lensing signal for pure NFW profiles in logarithmic annular bins to
estimate $\zetac$ using a given set of radii $(R_1, R_{o1}, R_{o2})$.
Thus, we use the signal for $R_1<R<R_{o1}$ to calculate the
first term in Eq.~\eqref{E:zetac} via direct summation over
the noisy mock data in broad logarithmic bins in $R$.  We
also estimate the second term using the fits to \dsr\ for
$R_1<R<4$\hmpc\ for the assumed value of \cvir. To do so, we
use the lensing profile for the best-fitting \mvir, determined to high
precision as in the start of Section~\ref{SS:theorresults}, and
estimate the second term using direct summation over the
numerically-determined (non-noisy) profile in narrow logarithmic bins
in $R$.  Given $\zetac$ estimated with and without the second term,
we then use our assumed $\cvir$ to convert the $\mproj(<R_1)$ to a 3d
virial radius \mvir, which (at fixed $\cvir$) is a simple
one-to-one mapping that can be determined via numerical integration.

In Table~\ref{T:nfwmap} we present the following, first without the
correction term for the outer annulus and then with it: the accuracy in recovering
$\mproj(<R_1)$, the accuracy in recovering $\mvir$, and the
statistical error on the recovered $\mvir$ relative to that from the
fit to \dsr\ using $R_1<R<4$\hmpc.  These results are shown for both
assumed concentration values, $\cvir=4$ and $7$, given the true 
profile with $\log_{10}[{h\mvir/\msun}] = 14.8$ and $\cvir=7$.
\begin{table*}
\begin{center}
\caption{Results of tests of NFW mass recovery for $\log_{10}[{h\mvir/\msun}] = 14.8$ and $\cvir=7$ when using the aperture mass statistic \zetac.\label{T:nfwmap}}
\begin{tabular}{ccccccccc}
\hline\hline
$R_1$ & $R_{o1}$ & $R_{o2}$ & $\mproj/M_\mathrm{2d,true}$ &
$\mvir/M_{200b,\text{true}}$ & $\sigma_M^{(\zetac)}/\sigma_M^{(\text{fit})}$ &
$\mproj/M_\mathrm{2d,true}$ &
$\mvir/M_{200b,\text{true}}$ & $\sigma_M^{(\zetac)}/\sigma_M^{(\text{fit})}$ \\
$\!\!\!$\hmpc$\!\!\!$ & $\!\!\!$\hmpc$\!\!\!$ & $\!\!$\hmpc$\!\!$ & \multicolumn{3}{c}{Neglect second term} &
\multicolumn{3}{c}{Estimate second term} \\
\hline
\multicolumn{9}{c}{Assume $\cvir=7$} \\
\hline
$0.275$ & $1.1$ & $2$ & $ 0.83$ & $ 0.74$ & $ 1.18$ & $ 1.00$ & $ 1.00$ & $ 1.40$ \\
$0.275$ & $1.1$ & $4$ & $ 0.83$ & $ 0.74$ & $ 1.18$ & $ 1.00$ & $ 1.00$ & $ 1.46$ \\
$0.275$ & $2$ & $4$ & $ 0.95$ & $ 0.90$ & $ 1.43$ & $ 1.00$ & $ 1.00$ & $ 1.54$ \\
$0.5$ & $1.1$ & $2$ & $ 0.66$ & $ 0.58$ & $ 0.87$ & $ 1.00$ & $ 1.00$ & $ 1.20$ \\
$0.5$ & $1.1$ & $4$ & $ 0.66$ & $ 0.58$ & $ 0.87$ & $ 1.00$ & $ 1.00$ & $ 1.24$ \\
$0.5$ & $2$ & $4$ & $ 0.90$ & $ 0.86$ & $ 1.12$ & $ 1.00$ & $ 1.00$ & $ 1.17$ \\
$1.1$ & $2$ & $4$ & $ 0.62$ & $ 0.55$ & $0.71$ &  $1.00$ & $ 1.00$ & $ 1.01$ \\
\hline
\multicolumn{9}{c}{Assume $\cvir=4$} \\
\hline
$0.275$ & $1.1$ & $2$ & $ 0.83$ & $1.26$ & $1.60$ & $1.01$ & $ 1.75$ & $ 1.95$ \\
$0.275$ & $1.1$ & $4$ & $ 0.83$ & $1.26$ & $1.60$ & $1.02$ & $ 1.80$ & $ 2.00$ \\
$0.275$ & $2$ & $4$ & $ 0.95$ & $1.62$ & $2.01$ & $ 1.02$ & $ 1.81$ & $ 2.24$ \\
$0.5$ & $1.1$ & $2$ & $ 0.66$ & $0.74$ & $0.95$ & $ 1.01$ & $ 1.35$ & $ 1.37$ \\
$0.5$ & $1.1$ & $4$ & $ 0.66$ & $0.74$ & $0.95$ & $1.02$ & $ 1.48$ & $ 1.44$ \\
$0.5$ & $2$ & $4$ & $ 0.90$ & $1.21$ & $1.15$ & $ 1.02$ & $ 1.50$ & $ 1.40$ \\
$1.1$ & $2$ & $4$ & $ 0.62$ & $0.72$ & $0.68$ & $ 1.04$ & $ 1.19$ & $ 1.08$ \\
\hline
\hline
\end{tabular}
\end{center}
\end{table*}

There are a few conclusions that can be drawn from this table.  First,
we begin with the idealized case in the top section of the table,
where the assumed $\cvir$ is the same as the true one.  In this case,
we see that depending on the configuration of the three radii used to
estimate \zetac, the projected mass may be underestimated by 5--40 per
cent if the second term in Eq.~\eqref{E:zetac} is ignored.  This
underestimate is propagated into an underestimate of the 3d \mvir\
that ranges from 10--45 per cent.  This underestimate due to ignoring
the mass in the outer annulus is less important
for $R_{o1}\gg R_1$ as it is for cases where the two radii are
relatively close to each other.  We also see that the statistical
error on the inferred \mvir\ from the aperture mass is typically
comparable to that for the fits to \ds\ using $R_1<R<4$\hmpc.

In this ideal case with the correct assumed \cvir, correcting for the
second term in \zetac\ using the best-fitting profile to \dsr\ for
$R_1<R<4$\hmpc\ leads to unbiased recovery of both $\mproj(<R_1)$ and
\mvir; however, the statistical errors on \mvir\ are larger than when
fitting to \dsr\ by typically tens of per cent.  This higher level of noise
is due to the noisy profile used to estimate the second term
in \zetac.  

Next, we consider the lower half of the table, in which we use a
profile with $\cvir=7$, and assume $\cvir=4$.  First, when we do not
include the second term in Eq.~\eqref{E:zetac}.  Second, when we 
include the second term in Eq.~\eqref{E:zetac}, the projected masses
are all slightly overestimated (by several per cent), and the 3d
\mvir\ are overestimated by 20--80 per cent (depending on $R_1$, with
smaller $R_1$ leading to larger biases).  We can explain the slight
overestimation of the \mproj\ when including the second term in
\zetac\ by the fact that we do the correction using profiles with a
low \cvir, which give too much mass in the outer regions from which the
second term is derived.  The significant overestimation of \mvir\
arises 
because, when we assume too low a concentration, then we anticipate a
profile with a low amount of mass on small scales, so the
conversion factor from $\mproj(<R_1)$ to \mvir\ is a large number.
This effect will be worse for small $R_1$, since the difference
between the lensing profiles for different concentrations is most
significant there.  If we allow a smaller variation, such as true
$\cvir=5$ and assumed $\cvir=4$, then we find a 10--20 per cent effect
on the 3d virial masses.

In general, the results for an input profile with $\cvir=4$ can be
understood as the inverse of the results given in
Table~\ref{T:nfwmap}.  However, for a less concentrated profile, the
bias in \mproj\ due to neglect of mass in the outer annulus is more
significant.  For a lower mass halo and fixed transverse
separation, the mass in the outer annulus is less important.  

We next consider the effect of centroiding errors on the aperture
mass.  When using the two mass models, we find that the projected
masses \mproj\ are systematically suppressed by 10--14 per cent due to
centroiding errors.  The exact level of suppression depends slightly but
not very strongly on the value of $R_1$ in the range we have
considered, and this suppression is then propagated into a suppression
of \mvir.

Because of the definition of \zetac, biases in the lensing signal
calibration that can be expressed as a single scale-dependent factor
enter linearly into the estimated masses in projection, $\mproj
\propto \ds$.  However, when using some model for the spherical
density profile to estimate the mass within some radius defined in
terms of a spherical over-density, such as $\mvir$, the mass will
scale even more strongly with \ds, because as the signal increases,
the spherical over-density radius moves outward, thus including more
mass in the total.  The exact scaling of the enclosed mass within some
spherical over-density depends on the model used to define the
appropriate radius, and on which over-density is used, but typically
the inferred $\mvir \propto \ds^{1.5}$.

One important point regarding the bias given in Table~\ref{T:nfwmap}
due to the wrong assumed concentration (for converting $\menc(<R_1)$
to \mvir) is that it 
has the same sign as the bias due to assumption of the wrong
concentration when fitting to \dsr.  Consequently, consistency of the
\mvir\ from the aperture mass calculation and the 
NFW fits to \dsr\ does not tell us whether the assumed concentration
is correct.  

In summary, we have found that the aperture mass
statistic \zetac\ has a strong dependence on the 
assumed \cvir\ when converting the extracted projected masses to 3d
\mvir.  An additional problem is that a (much less
concentration-dependent) correction must be used to properly correct for the
term from the outer annulus $R_{o1}<R<R_{o2}$; otherwise, the
projected masses can be underestimated by tens of per cent, an effect
that is worse for more massive clusters.  While less affected by
centroiding errors than fits to \dsr\ that use scales below
$0.5$\hmpc, the aperture mass statistic can still be suppressed by
roughly ten per cent due to centroiding errors (or any of the other
errors from Section~\ref{SS:challenges-obs} that have a similar form).   Finally, it can be
substantially noisier, typically by 50 per cent, than fits to \dsr\
using the same scales (which means that it is noisier than fits to \ups).

In principle, these biases due to the concentration-dependence of the
2d to 3d converstion may be removed if the conversion from
$\mproj(<R_1)$ to \mvir\ is carried out using the best-fitting NFW
profile from fits to both \cvir\ and \mvir, as in 
\cite{2009arXiv0903.1103O}.  However, as will be shown in the next
section, these fits tend to be substantially noisier due to the
additional fit parameter, which will further amplify the noise on
the recovered mass from \zetac.  Thus, this approach is not very
advantageous relative to the fits to \ups, which are similarly
insensitive to the assumed concentration but are only slightly noisier
than fits to \dsr.
\begin{figure}
\begin{center}
\includegraphics[width=\columnwidth,angle=0]{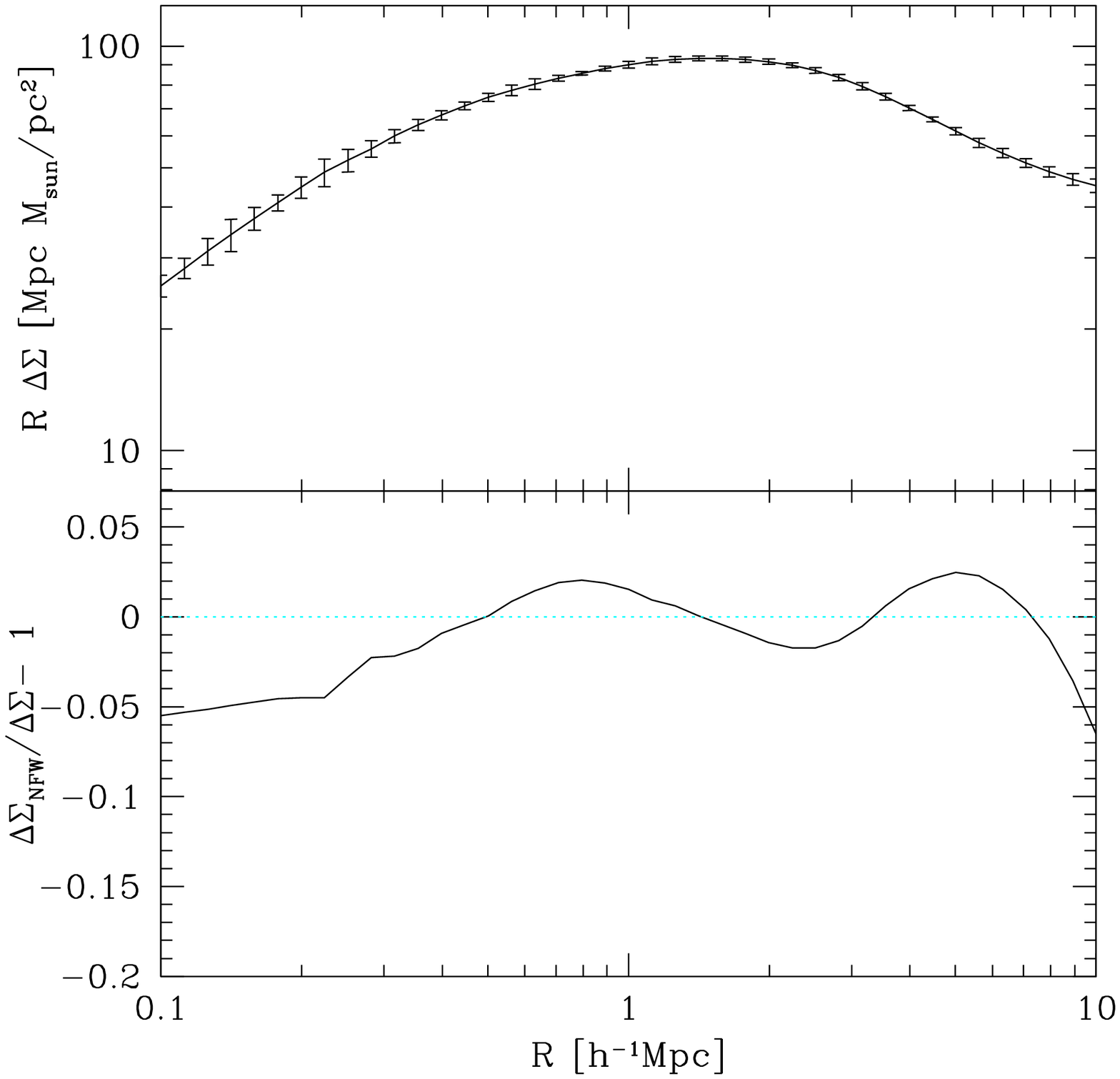}
\caption{\label{F:simsignal} {\em Top:} Lensing signal $R \dsr$ from
  simulations for the higher mass (lower number density) threshold sample described in the text.
  The solid lines with error-bars show the signal stitched to an NFW
  profile with $\cvir=5$ for $r<0.2$\hmpc\ (to remove resolution
  effects).  {\em Bottom:} Ratio of the
  signal for the best-fitting NFW profile to the true simulation
  signal.}
\end{center}
\end{figure}

\subsection{Profiles from $N$-body simulations}\label{SS:simresults}

In this section, we present the results of tests of mass estimation
using cluster profiles measured from the simulations described in
Section~\ref{S:simulations}.  The properties of these simulated
cluster samples are summarized in Table~\ref{T:testprofiles}.  We use the signal from simulations for
mass threshold samples selected by taking all clusters above some
\mvir\ such that $\overline{n}=0.25$ $2$, and $16\times 10^{-6}
(h/\mathrm{Mpc})^3$, with the first of these samples shown in Fig.~\ref{F:simsignal}.
The samples have mean masses $\langle\mvir\rangle=7.36$, 
$3.95$, and $1.55\times 10^{14}$\hmsun, though the stitching to NFW profiles below
certain scales as described in Section~\ref{S:simulations} increases
the mass by several per cent.  All comparisons between estimated
\mest\ and true \mtrue\ take this small increase into account.  The
error-bars shown in Fig.~\ref{F:simsignal}, which include cosmic
variance, are estimated by dividing the eight simulation boxes each
into 20 sub-volumes comparable in size to that of the maxBCG cluster
sample, and finding the variance of the signal between the 160 total
sub-volumes.  We have only shown the case of stitching to NFW profiles
with $\cvir=5$ at 0.2\hmpc\ in Fig.~\ref{F:simsignal}; when stitching
to an NFW profile with $\cvir=7$ at 1\hmpc, the signal on smaller
scales is steeper.  In the former case, this resolution correction
increases  the mass by 1.5
per cent compared to the mass in the simulations; in the latter case,
the correction is 6 per cent.

In the bottom panel of Fig.~\ref{F:simsignal}, we compare \dsr\ from
the simulations to that for the best-fitting NFW profile (determined
by varying both \mvir\ and \cvir\ and fitting using $0.2<R<2$\hmpc).
As shown, for most of the scales of interest, the deviations are less
than 5 per cent.  We see that the NFW profile
overestimates the signal on $\sim 3$--$8$ \hmpc\ scales.  This result
is consistent with that from \cite{2004MNRAS.350.1038C}, who also find
that on large scales the density profiles fall off faster than NFW.
The effect is more significant when expressed in terms of the density
profile $\rho(r)$.  On the largest scales shown here, as $R$
approaches $10$\hmpc, the NFW profile signal starts to be too low,
because the simulation includes contributions from LSS (again, this
effect is more pronounced in $\rho(r)$ and appears at lower radii).

For the subsections that follow, we have added realistic levels of
shape noise to the signal, based on calculations of the lensing signal
using the maxBCG cluster catalogue with similar number density samples.

\subsubsection{Parametric modeling}\label{SSS:ressimpara}

\begin{figure*}
\begin{center}
\includegraphics[width=5in,angle=0]{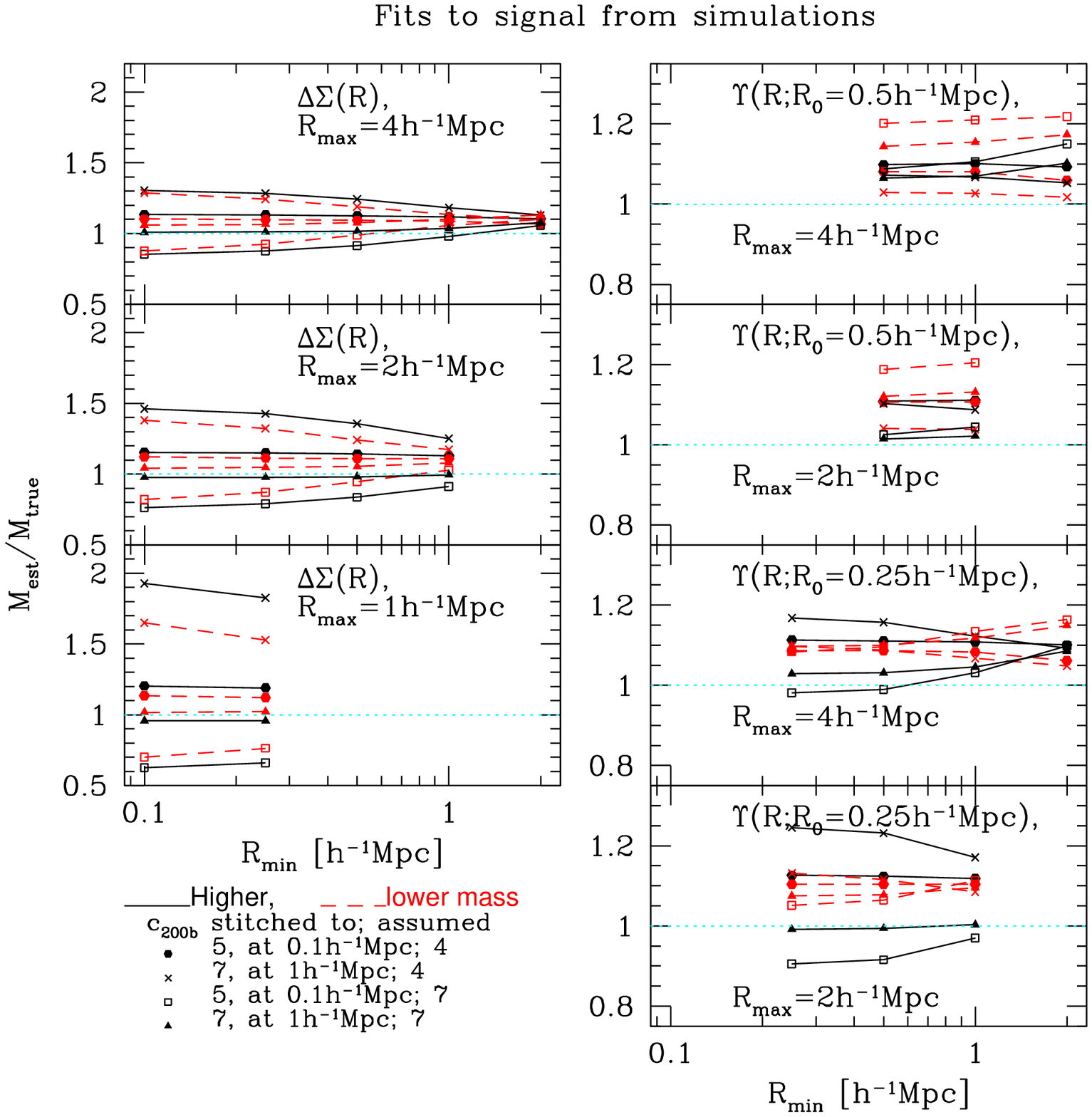}
\caption{\label{F:showsimfit}
Results for $\mest/\mtrue$ as a
function of the minimum fit radius, \rmin, from parametric fits to the lensing signal from simulations, for
seven different combinations of observable (\dsr\ or \ups) and
\rmax\ shown separately in each panel.  As indicated in the legend,
line colours and types are used to indicate the mass scale, whereas
point styles are used to indicate the input signal (which NFW profile
was used to correct for resolution effects) and assumed concentration
in the fits, either $\cvir=4$ or $7$.  The horizontal dotted line on
each panel shows the ideal unbiased result.  The vertical axis is the
same for all panels in the left column, and spans a smaller range for
all panels in the right column so that the details will be more visible.}
\end{center}
\end{figure*}

We begin by showing the effects of parametric modeling of the lensing
profiles from simulations.  We use the three aforementioned mass
threshold samples, with the two methods of connecting to NFW profiles
(Section~\ref{S:simulations}) to correct for resolution effects:
$\cvir=5$ at $r=0.2$\hmpc, and $\cvir=7$ at $r=1$\hmpc.  We then fit
to \dsr\ and \ups\ with $R_0=0.25$ and $5$\hmpc, with varying \rmin\ and \rmax,
for our two extreme concentration values of $\cvir=4$ and $\cvir=7$.
The fitting procedure is the same as for the analytic profiles
  in Section~\ref{SSS:analytic-para}.
Fig.~\ref{F:showsimfit} shows the results of these fits for the
highest and lowest of the mass threshold samples.

The important point to consider in this plot is that we would like the
output mass from a given estimator to be relatively insensitive to the
form of the inner profile (represented by the two different
connections to NFW profiles on small scales) and to the assumed
concentration.  Furthermore, we would like it to be only weakly
dependent on the mass, assuming that corrections for systematic bias
will be derived from simulations, but that strong mass dependence may
be difficult to calibrate out correctly.  Consequently, what we hope
to see in an optimal estimator of cluster mass is that all the lines
on a given panel (representing the results with different input
profiles, assumed concentrations, and masses) give very similar
results; we do not want to use an estimator that has large scatter
between the lines.  So, for example, the lower left panel shows, as we
already saw with pure NFW profiles in Section~\ref{SS:theorresults},
that fitting \dsr\ to NFW profiles with $\rmax=1$\hmpc\ and a fixed
concentration leads to very large systematic uncertainties, more than
a factor of 2 total range in the best-fit masses.  As we increase
\rmax, we become less sensitive to the inner details of the profile,
so the scatter between the lines becomes less significant, but for
$\rmin\le 1$\hmpc\ they still cover a range of $\sim 40$ per cent in
mass even for $\rmax=4$\hmpc, well outside the virial radius.  For
$\rmin=2$ and $\rmax=4$\hmpc, the systematic uncertainty is only $\sim
10$ per cent; however, the statistical error on the mass (not shown on
this plot) has roughly doubled relative to the results with $\rmin\le
0.5$\hmpc.

In contrast, we see that $\Upsilon(R;R_0=0.25$ and $0.5$\hmpc), in the
right panels in Fig.~\ref{F:showsimfit}, performs quite well.  The
difference between the two mass threshold samples suggests that a
larger $R_0\sim\rmin$ is preferable for samples with larger halo
masses, with minimal profile-related systematics for $R_0=0.5$\hmpc\
for the sample with a mass above $7\times 10^{14}\hmsun$, and
$R_0=0.25$\hmpc\ for the sample with a mass around $1.6\times
10^{14}$\hmsun (and therefore smaller scale and virial radii).  While
the cluster mass is not known a priori, a preliminary fit with one
choice of $R_0$ could be used to estimate an approximate mass, and
then a new $R_0$ could be chosen to be around $1/4$ to $1/5$ of the
virial radius, provided that this scale is reliable from the
perspective of small-scale systematics (Section~\ref{SS:challenges-obs}).

In all cases, \ups\ does not converge to the
true mean mass, for two reasons: (1) the lensing signal includes a
small but non-negligible contribution due to large-scale structure on
the scales we have used, leading to an overestimation of \mest; and
(2), even on scales where LSS is not important, the simulation
profiles fall off faster than the NFW model, which somewhat
counteracts the previous effect.  Fortunately, since it is relatively
insensitive to the inner details of the profile, the assumed
concentration, and the mass, this systematic positive bias in the
masses can be calibrated out using simulations, whereas systematic
uncertainty in \dsr-based mass estimates due to concentration
assumptions and small-scale effects cannot be calibrated out in this
way.

Some differences in these results from Section~\ref{SS:theorresults}
can be attributed to the LSS in the simulations that was not put into
the pure NFW profiles, and to the fact that the simulation profiles
are not strictly NFW profiles.  So, for example, in
Fig.~\ref{F:theorresults}, the results for fitting to \dsr\ converge
to the true mass on large scales if the right concentration is
assumed, whereas the fitting to \dsr\ in simulations converges to a
mass that is too high by 5 to 10 per cent when using the largest
scales only.

As in Section~\ref{SS:theorresults}, we point out that for a stacked
cluster sample, the level of variation we have allowed in the assumed \cvir\
is likely excessive from the standpoint of $N$-body simulations.
However, given the systematic profile changes that may occur due to
baryonic effects, centroiding errors, and intrinsic alignments, the
variation we have assumed is not entirely unreasonable.  For fits to
individual cluster lensing data, the variation we have assumed is
quite reasonable, and possibly even an underestimate of the true
variation, given the large lognormal scatter in cluster concentrations
in $N$-body simulation plus these other systematics that change the
profile on small scales.

We also estimate the effects of centroiding errors on the parametric
mass recovery.  As for the theoretical profiles, we use the model for
centroiding errors given in \cite{2007arXiv0709.1159J}, with 
offset fractions of 20 and 25 per cent for the lower and higher
abundance thresholds, respectively.

Here we describe how centroiding errors modify the curves that
were shown in 
Fig.~\ref{F:showsimfit}.  As we have seen before, the
offsets suppress masses estimated directly from \dsr, with larger
biases when restricting to smaller scales.  Furthermore, the profiles
with more mass in the inner regions are more strongly affected.  For
example, the simulation signal stitched to NFW with $\cvir=7$ at $1$\hmpc\ is
more strongly affected than the signal stitched to $\cvir=5$ at
$0.2$\hmpc.  Given that the former resulted in mass estimates that were above
the masses estimated from the latter 
when fitting to \dsr\ (without offsets,
Fig.~\ref{F:showsimfit}) by up to tens of per cent depending on the
value of \rmax, the net effect of offsets is to lower all estimated
masses while also reducing the difference between the curves, since
those with the two stitched profiles now tend to agree more closely.
For example, when using $\rmax=1$\hmpc, the values of $\mest/\mtrue$
without including centroiding errors in the modeling range from $0.6$
to $1.9$ (factor of three). Centroiding errors in the input
data reduce the range of $\mest/\mtrue$ to $0.4$ to $0.9$ (factor of
two), where the main cause of this variation is the assumed value of
\cvir\ rather than the input profile.  For $\rmin=0.5$ and
$\rmax=4$\hmpc, $\mest/\mtrue$ ranges from $0.9$ to $1.25$ when we do
not include centroiding errors, whereas when we include them, it
ranges from $0.8$ to $1$.  As we have seen before when using pure NFW
profiles, \ups\ with $\rmin=R_0=0.5$\hmpc\ is almost completely
insensitive to this model for centroiding errors when using
$\rmax=4$\hmpc\ (masses are suppressed at the 10 per cent level with
$\rmax=2$\hmpc).  This insensitivity to such systematics makes the
ADSD statistic \ups\ the optimum choice for
parametric mass fitting on stacked clusters selected from imaging
data, which is prone to centroiding errors of this variety.

In principle, explicit modeling of the offset distribution, as in
\cite{2007arXiv0709.1159J}, can remove its effects when fitting to
\dsr.  However, the exact results may be sensitive to the details of
the centroiding model used and its accuracy when compared to the true
distribution, which is not typically well known.  For example, that
paper uses mock simulations to estimate the centroiding error
distribution, which means that this model is quite sensitive to the
realism of the model for populating the simulation dark matter halos
with galaxies.  Furthermore, the other systematic uncertainties
associated with using \dsr\ (e.g., sensitivity to baryonic effects and
intrinsic alignments) remain, whereas their influence on \ups\ is much
smaller.

\begin{figure*}
\begin{center}
\includegraphics[width=5in,angle=0]{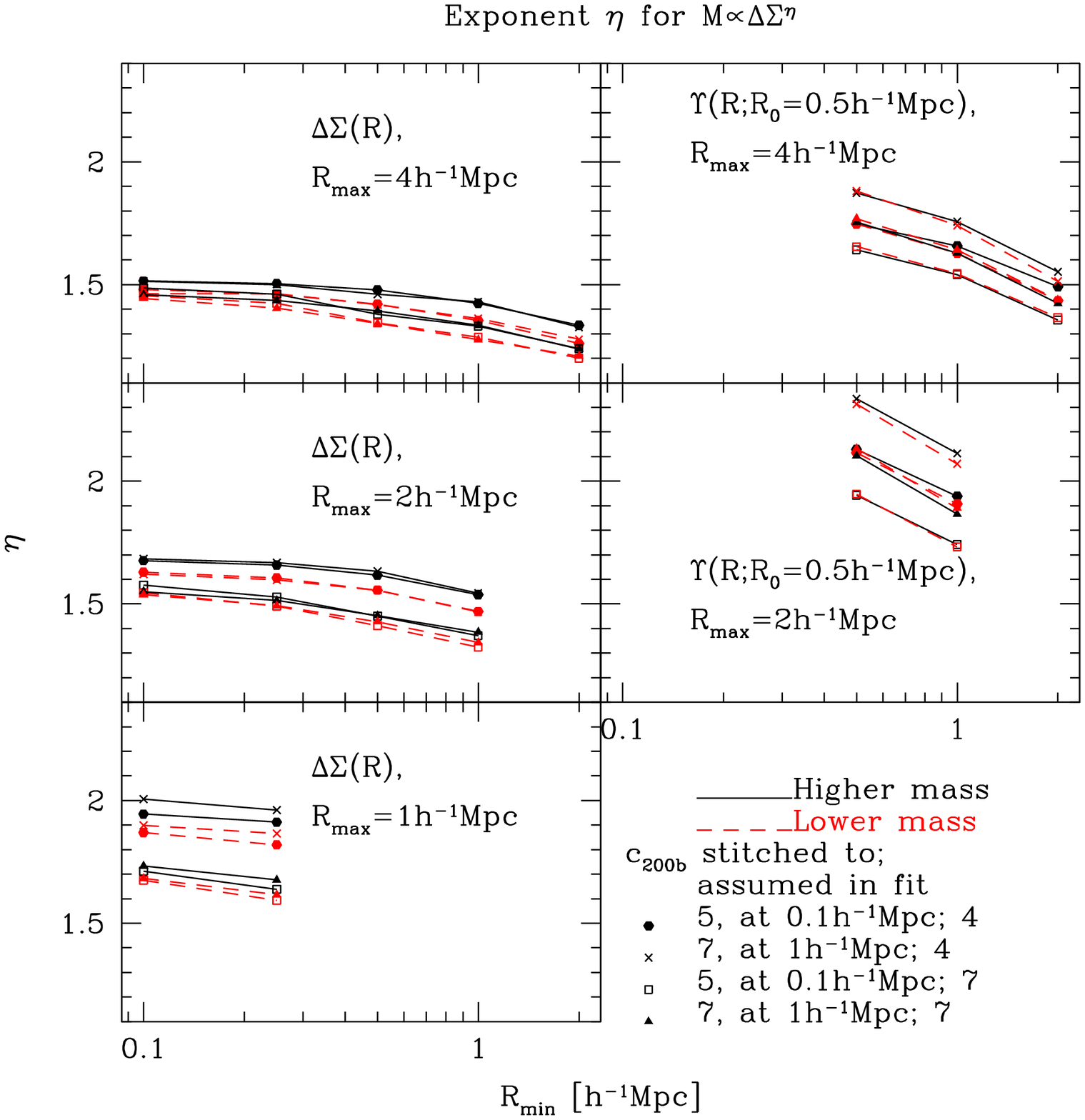}
\caption{\label{F:calibbias} Results for $\eta$, the scaling of the
  estimated \mvir\ with the lensing signal calibration, as a function
  of the minimum fit radius, \rmin, from parametric fits to the
  lensing signal from simulations, for five different combinations of
  observable (\dsr\ or \ups) and \rmax\ shown separately in each
  panel.  We only show $\eta$ for the highest and lowest mass threshold sample.}
\end{center}
\end{figure*}

Another issue we consider is the effect of overall lensing signal
calibration biases on the estimated masses.  As a test, we use the
signals from simulations multiplied by factors of $0.9$ and $1.1$, and
refit for the masses.  The results are used to estimate a power-law
relation $\mvir \propto \ds^\eta$, and $\eta$ is determined for
the different mass scales, stitched signals, assumed concentrations,
fit method (\dsr\ or \ups), and minimum and maximum fit radii. Note
that $\eta$ is also dependent on the spherical over-density used to
define the profile, though we do not explore this effect in detail.
A naive scaling of surface mass density with mass predicts $\eta=1.5$, 
but other effects will modify this. 
The results of this test are shown in Fig.~\ref{F:calibbias}.

As shown, $\eta$ is a decreasing function of \rmin\ and \rmax.  When
fitting to \dsr, $\eta$ does not depend on the details of the inner
profile, and is larger for higher masses and lower assumed \cvir, with
the dependence on \cvir\ being the more significant dependence.  For
example, when fitting to \dsr\ for the lower mass sample from
simulations stitched to an NFW profile with $\cvir=5$ at $0.2$\hmpc,
using $0.5<R<4$\hmpc\ for the fits, we find that $\mvir \propto
\ds^{1.42}$.  In contrast, when fitting to \ups\ with $R_0=0.5$\hmpc,
the trends in the fitting mass with calibration are stronger for a
given combination of $(\rmin,\rmax)$.  Here, we see that there is
minimal dependence on mass, and some small dependence on the details
of the profile and the assumed concentration.  For the same case
considered when fitting to \dsr, we find $\eta=1.75$, an increase of
23 per cent.  Consequently, systematic errors in \ups\ due to
miscalibration of the lensing signal are larger than systematic errors
in \dsr\ (assuming that other aspects of the fit, such as \rmin\ and
\rmax, are similar).

Next, we briefly discuss the effects of allowing both \cvir\ and \mvir\
to vary, rather than fixing \cvir, as in \cite{2009arXiv0903.1103O}.
While this procedure has the disadvantage of increasing the
statistical errors on the mass, it does allow for improved mass
recovery.  Our results suggest that with NFW fits to \dsr\ with
$0.5<R<4$\hmpc, the degeneracy between \mvir\ and \cvir\ is such that
$\mvir\propto\cvir^{-1/3}$.  This result explains the magnitude of the
deviations from the true mass when the concentration is fixed to a
value that is not consistent with the best-fitting concentration
(though, again, the deviations in concentration we have tested are not
sufficiently bad that the fit $\chi^2$ values reveal a clear
discrepancy).  In contrast, the exponent on that scaling between
\mvir\ and \cvir\ is far closer to zero when fitting to \ups\ with
$R_0=0.5$\hmpc\ using the same scales: $\mvir\propto\cvir^{0.05}$.
This degeneracy becomes more striking when the fits are restricted to
smaller scales, e.g., $\mvir\propto\cvir^{-1}$ when fitting to \dsr\
using $0.1<R<1$\hmpc.

When fitting the simulation signals with both \mvir\ and \cvir\ as
free parameters, we find that even when centroiding errors are
included in the data, the fits are able to recover the masses for both
mass scales and inner profiles, for several types of fits that we
attempted (using \dsr\ from $0.5<R<4$\hmpc, from $0.1<R<1$\hmpc, and
using \ups\ with $R_0=0.5$ and $0.5<R<4$\hmpc).  When using data from
$0.5<R<4$\hmpc, the deviations of the signal in simulations from an
NFW profile led to best-fitting masses that are 5 per cent higher than
the true masses; when fitting from $0.1<R<1$\hmpc, the best-fitting
masses are only $\sim 2$ per cent higher than the true ones, because
the deviations from NFW are not as striking on those scales, and the
large-scale structure term is also more negligible.  The mass
estimates tend to be noisier in this case, and the concentrations that
are recovered are highly suppressed relative to the true
concentrations when centroiding errors are included (e.g., from
best-fitting $\cvir\sim 5$ and $\sim 6.5$ without centroiding errors,
down to $3$ and $3.5$ with centroiding errors).  We find that these
two-parameter fits for mass and concentration lead to statistical
errors in the masses that are larger than the errors in one-parameter
fits by approximately 45 per cent.  This increase is larger than 
the increase when fitting to \ups\ (14 or 32 per cent, for $R_0=0.25$
or $0.5$\hmpc), and \ups\ has the additional advantage
of removing the impact of small-scale systematics, which would still
be present when fitting \dsr\ to the two-parameter model.

Finally, when fitting with free \mvir\ and \cvir, the dependence of
the best-fitting masses on the lensing signal calibration is reduced.
For example, when fitting to \dsr\ using $0.5<R<4$\hmpc\ and fixed
concentration, we had found previously that $\mvir\propto\ds^{1.42}$.
When the concentration is allowed to vary, that exponent becomes
$\eta=1.25$.  This change results from the fact that if the signal
increases, the assumed mass and therefore $\rvir$ increases as well,
so for a fixed scale radius determined from the data, the
concentration would naturally tend to increase.  When fitting to \ups,
$\eta$ is not affected by whether or not the concentration is fixed.

In summary, we have found that \ups\ is the optimal statistic for
parametric mass modeling given its insensitivity to the profile at
small scales, with $R_0=0.25$--$0.5$\hmpc\ for the cluster masses used here, 
giving a reasonable
compromise in reducing systematic error while retaining reasonable
$S/N$ on the recovered masses for the case discussed here, but in
general the choice of $R_0$ will depend on the specific application
one has in mind and on the scales to which the data can be considered
relatively systematics-free.  This statistic tends to slightly
overestimate the mass due to the combination of two competing effects:
the profile deviation from NFW on large scales, and the neglected
large-scale structure contribution to the lensing signal.  However,
these effects are only very weakly dependent on the details of the
profile, the mass, and the cosmology, making them easy to calibrate
out at the few per cent level using $N$-body simulations.  This result
is in stark contrast to the effect of small-scale systematics on the
masses estimated from \dsr\ (e.g., varying concentrations, and
deviations from NFW due to intrinsic alignments and baryonic effects),
which lead to larger systematic uncertainties in the
recovered masses.  These conclusions hold in cases where the NFW
concentration is fixed.  If it is allowed to vary, then the
statistical errors will increase more than when using \ups\ with a
reasonable $R_0$, but systematic errors decrease, provided that the
systematic errors in the lensing signal appear reasonably similar to a
change in NFW concentration, which is not the case for several of the
small-scale systematics in Section~\ref{SS:challenges-obs}.

\subsection{Example application with data}\label{SS:exampledata}

Here we consider the maxBCG cluster lensing data in six scaled
richness bins ($12\le\nt\le 79$), which was previously used in
\cite{2008JCAP...08..006M} for joint estimation of the
concentration-mass relation and the mass-richness
relation.  Here, we use several examples of
fixed concentration-mass relations and several of the fitting methods
considered in the previous sections to estimate the mass-richness
relation, always with $\omegam=0.25$.  This estimation as follows:

\begin{itemize}
\item We generate 200 bootstrap-resampled datasets to estimate
  the noise in the data.  For this bootstrap procedure, the data are divided
  into 200 regions on the sky which are bootstrapped (rather
  than bootstrapping the individual lenses).  More details on this
  procedure is given in \cite{2008JCAP...08..006M}.
\item For each dataset, we separately fit the data in each
  richness bin for \mvir\ assuming some $\cvir(\mvir)$
  relation and fit method, for each richness bin.  The choice
  of fit method includes specifying the statistic to fit and the range
  of transverse separations to use.  Thus, given
  logarithmic bins in transverse separation denoted $i$ ($R_i$), dataset $j$,  richness bin
  $k$, statistic for a given fit method $\ell$ (denoted $\Xi$ for $\Xi=\ds$ or \ups),
  and $\cvir(\mvir)$ relation $m$,
  we use the Levenberg-Marquardt algorithm to separately minimize the $j\times k\times \ell\times m$ values of $\chi^2$ defined as
  follows:
\beq
\chi_{jk\ell m}^2 = \sum_i
  \frac{(\Xi^{\text{(data)}}_{jk\ell}(R_i)-\Xi^{\text{(model)}}_{\ell
  m}(R_i))^2}{\sigma^2(\Xi_{k\ell}(R_i))}
\eeq
where we use $i$ such that $R_{\text{min},\ell}\le
  R_i\le R_{\text{max},\ell}$.  The result of this
  procedure is a matrix with $j\times k\times\ell\times m$ values of
  \mvirt, where in practice we use $j=200$, $k=6$, $\ell=3$, and
  $m=3$.  The fit methods and concentration-mass relations are
  described in detail below. 
\item The set of $k$ \mvir\ values for a given dataset ($j$), fit
  method ($\ell$), and concentration-mass relation ($m$) are used to fit for a power-law
  relation between scaled richness and halo mass:
\begin{equation}\label{E:powlawmn}
\mvir^{\text{(model)}} = \left[\left(\mvirt \times 10^{14}\right)\hmsun\right]\left(\frac{\nt}{20}\right)^\gamma
\end{equation}
This fit has two parameters: an amplitude \mvirt\ that is the mass 
at our pivot richness of $\nt=20$ in units of $10^{14}$\hmsun, and an
exponent $\gamma$.  We find the best-fitting values of \mvirt\ and
$\gamma$ for each $(j,\ell,m)$ by minimizing
\beq
\chi_{j\ell m}^2 = \sum_k \frac{(M_{200b,jk\ell
    m}^{\text{(data)}}-\mvir^{\text{(model})}(N_{200,k}))^2}{\sigma^2(M_{200b,\ell m})}.
\eeq
The result is a matrix with $j\times \ell\times m$ values of \mvirt\
and $\gamma$.
\item We use the list of $j$ power-law fits for each bootstrap-resampled dataset
  to estimate the mean and variance of \mvirt\ and $\gamma$ for a
  given combination of fit method $\ell$ and concentration-mass
  relation $m$.
\end{itemize}

We include $m=3$ concentration-mass relations in our tests: a power law
with 
\beq\label{E:powlawcm} 
\cvir = 5\left(\frac{\mvir}{10^{14}\hmsun}\right)^{-0.1}, 
\eeq 
consistent with
\cite{2008JCAP...08..006M}; a constant $\cvir=4$; and a constant
$\cvir=7$.  We examine the results for $\ell=3$ fit methods: an extreme
one assuming a small aperture for the cluster data, using \dsr\ from
$0.1$--$1$\hmpc; using \dsr\ from $0.5$--$4$\hmpc; and using \ups\
with $R_0=0.5$\hmpc\ from $0.5$--$4$\hmpc, given its good performance
on theoretical profiles and simulations in the previous sections.
We consider the fit results first without and then with correction factors derived from
Fig.~\ref{F:showsimfit}.  While we only have
simulation-based correction factors for
samples with three mean masses (which are threshold samples, not
discrete mass bins as in this example) and two concentrations, we
interpolate those results to derive approximate corrections for all
the fits done in this section.

The final type of correction that we apply is a calibration factor that
reduces the lensing signal calibration from \cite{2008JCAP...08..006M}
and \cite{2008MNRAS.390.1157R} by 6 per cent.  The reason for this
correction is that for 30 per cent of the spectroscopic training set presented in
\cite{2008MNRAS.386..781M} for calibration of photometric
redshifts that are used to estimate the lensing signal, an incorrect
photometric calibration was used when computing the photometric
redshifts.  We emphasise that this incorrect
calibration was only used for the {\sc kphotoz} photometric redshifts,
not for any other photometric redshift sample, and thus the
lensing signal calibrations that are quoted for other photometric redshift methods in
that paper are correct.  As a result of this error, the calibrations from {\sc kphotoz}
which were used for the data in \cite{2008JCAP...08..006M}
and \cite{2008MNRAS.390.1157R} that we analyse here were 6 per cent too
high, so we now apply a correction to the signal.  We then present
the results for the best-fitting masses after application of both the
signal calibration correction and the simulation-based correction
factors due to the mass estimation method.  

\begin{figure*}
\begin{center}
\includegraphics[width=5in,angle=0]{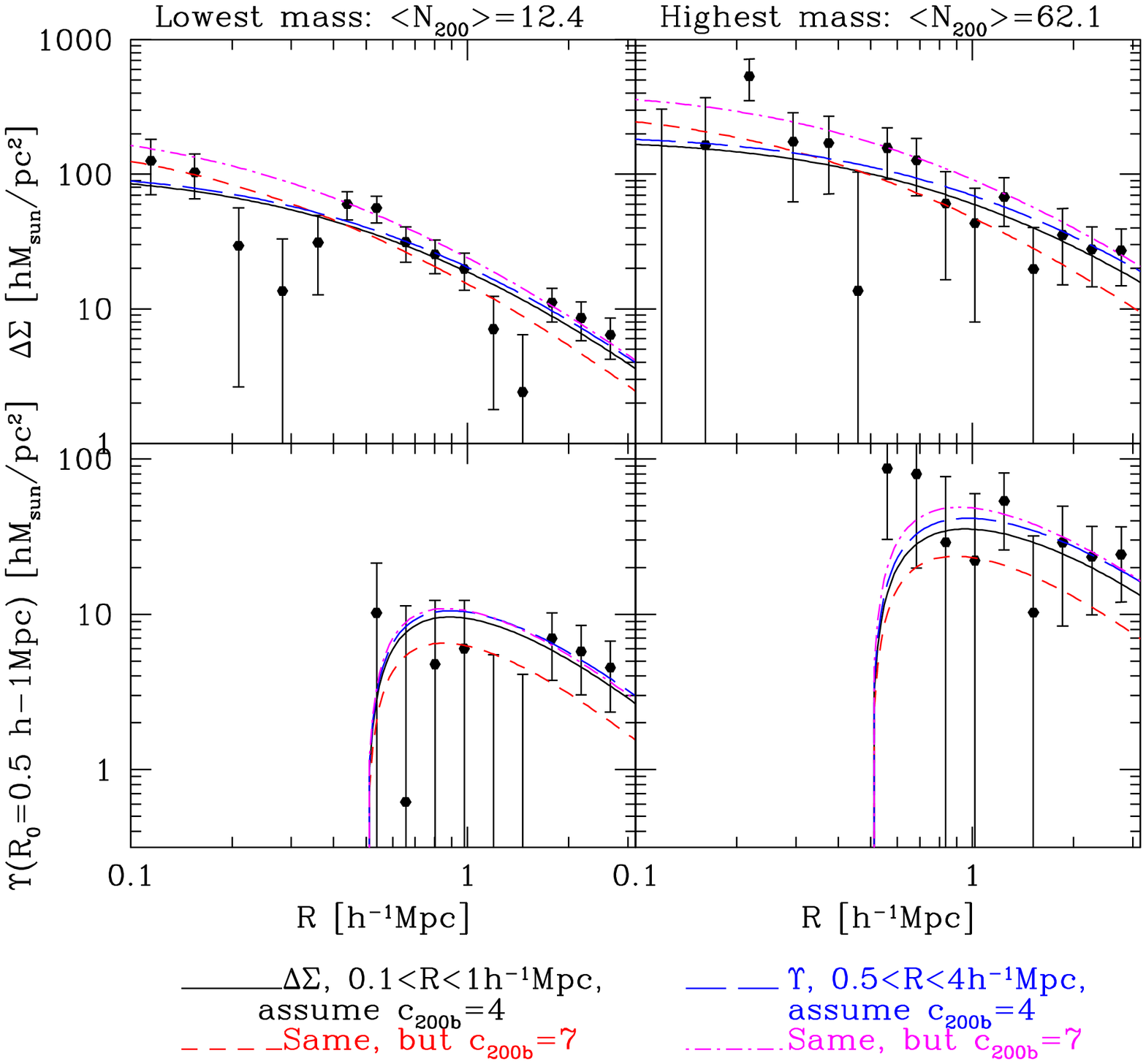}
\caption{\label{F:examplesig} Observed lensing signal from
  \protect\cite{2008JCAP...08..006M} for stacked maxBCG clusters, presented as
  \dsr\  and \ups\ with $R_0=0.5$\hmpc\ in the top and bottom panels,
  respectively.  We show the lowest (left) and highest (right) richness bins out of
  the six used for the analysis.  In addition to the data with
  bootstrap error-bars, we also show four theoretical signals labelled
  on the plot.  Two of them were derived by fitting \dsr\ using
  $\rmin=0.1$ and $\rmax=1$\hmpc\ with different assumed
  concentrations; the other two, by fitting \ups\ with $R_0=\rmin=0.5$
  and $\rmax=4$\hmpc.  The 6 per cent calibration correction described
  in the text has been applied.  Because we required a power-law
  relationship between mass and richness, the best-fitting signals
  shown for these two bins were influenced by the data in the other
  richness bins (not shown).}
\end{center}
\end{figure*}

Fig.~\ref{F:examplesig} shows the observed signal for the lowest and
highest richness bins for $0.1<R<4$\hmpc, and the theoretical signal
from the fits.  This theoretical signal is derived by taking the
best-fitting mass-richness relation, evaluating it at the mean
richness of the bins that are shown, and using the resulting mass and
assumed concentration to define the theoretical signal.  The fits did
not only use the data shown on the plot, because the requirement of a
power-law mass-richness relation means that the theoretical signal at
the richness bins shown was also influenced by the data in all other
bins.

For reference, given a best-fitting mass-richness relation from
Eq.~\eqref{E:powlawmn} with $\mvirt=1.55$ and $\gamma=1.15$ (which is a
typical value given the scatter between the results in
Table~\ref{T:powlawmn}), the combination with Eq.~\eqref{E:powlawcm}
gives a concentration-richness relation of 
\beq\label{E:powlawcn}
\cvir = 4.78 \left( \frac{\nt}{20} \right)^{-0.115} 
\eeq 
Thus, within our richness range of $12\le\nt\le79$, the concentrations
vary from $5$ to $4$ as we move from the lowest to the highest
richnesses. When we instead fix $\cvir=4$ independent of mass, we
lower the concentrations at the low \nt\ end of the sample by 20 per
cent, without changing the concentrations at the very high mass end.
When we fix $\cvir=7$, then we raise all the concentrations by a very
significant amount, from $\sim 40$ per cent increases at the low mass
end to $75$ per cent at the high mass end.  The results for the three
concentration-mass relations and fitting methods are given in
Table~\ref{T:powlawmn}.

\begin{table*}
\begin{center}
\caption{Results of power-law fits for a mass-richness relation using
  stacked maxBCG cluster lensing data, using
  three fit methods and three concentration-mass relations.  First
  present the best-fitting masses; then, include corrections for the
  bias on the mass estimation from simulations
  (Fig.~\ref{F:showsimfit}); finally, with both the simulation
  corrections and a 6 per cent decrease of the amplitude on the
  lensing signal, as described in the text.\label{T:powlawmn}}
\begin{tabular}{cccccccc}
\hline\hline
Fit method & $\!\!\!\!\cvir(\mvir)\!\!\!\!$ & \mvirt\ & $\gamma$ &
\mvirt\ & $\gamma$ & \mvirt\ & $\gamma$ \\
 & & \multicolumn{2}{c}{No correction} & \multicolumn{2}{c}{Simulation
   correction} & \multicolumn{2}{c}{Sim. and photo-$z$ corrections} \\
\hline
\dsr,  & Eq.~\eqref{E:powlawcm} & $1.64\pm 0.20$ & $1.24\pm 0.35$  &
1.31 & 1.10 & $1.19\pm 0.10$ & $1.10\pm 0.28$ \\ 
$0.1<R<1$\hmpc$\!\!\!\!$ & $\cvir=4$ & $1.58\pm 0.15$ & $1.07\pm 0.26$
& 1.14 & 1.01 & $1.04\pm 0.09$ & $1.01\pm 0.24$ \\
 & $\cvir=7$ & $1.01\pm 0.08$ & $0.93\pm 0.22$ & 1.16 & 0.98 &
 $1.06\pm 0.09$ & $0.98\pm 0.24$ \\
\hline
\dsr, & Eq.~\eqref{E:powlawcm} & $1.72\pm 0.13$ & $1.18\pm 0.18$ & 1.56
& 1.14 & $1.44\pm 0.10$ & $1.14\pm 0.17$ \\
 $0.5<R<4$\hmpc$\!\!\!\!$ & $\cvir=4$ & $1.70\pm 0.12$ & $1.14\pm
 0.16$ & 1.51 & 1.11 & $1.40\pm 0.10$ & $1.10\pm 0.16$ \\
 & $\cvir=7$ & $1.52\pm 0.10$ & $1.06\pm 0.15$ & 1.46 & 1.11 &
 $1.35\pm 0.09$ & $1.11\pm 0.16$ \\
\hline
\ups, & Eq.~\eqref{E:powlawcm} & $1.79\pm 0.18$ & $1.20\pm 0.24$ & 1.67
& 1.20 & $1.50\pm 0.16$ & $1.21\pm 0.24$ \\
 $R_0=0.5$\hmpc,$\!\!\!\!$ & $\cvir=4$ & $1.81\pm 0.18$ & $1.18\pm
 0.23$ & 1.75 & 1.16 & $1.56\pm 0.16$ & $1.17\pm 0.23$ \\
$0.5<R<4$\hmpc$\!\!\!\!$  & $\cvir=7$ & $2.02\pm 0.19$ & $1.11\pm
0.21$ & 1.73 & 1.17 & $1.56\pm 0.16$ & $1.17\pm 0.23$ \\
\hline
\end{tabular}
\end{center}
\end{table*}

We begin by discussing the first fit method, using \dsr\ from
$0.1<R<1$\hmpc.  As we have noticed in previous examples with the
theoretical profiles and simulations, the results using these scales
are highly sensitive to the assumed concentration-mass relation.  We
see that changing the assumed concentration among our three options
leads to 50 per cent variation of the amplitude \mvirt, significantly
larger than the statistical errors on this parameter, when we do not
impose corrections from the simulations.  The exponent
$\gamma$ undergoes 20 per cent changes, which are roughly consistent
with the size of the $1\sigma$ statistical errors.  The changes in
this exponent can be easily understood as follows.  First, if we
change from the power-law concentration-mass relation in
Eq.~\eqref{E:powlawcm}, to fixed $\cvir=4$, then we are lowering the
assumed concentration for all but the highest mass halos.  This means
that, due to the typical concentration-mass anti-correlation when
fitting \ds, the best-fitting masses should increase at the lower mass
end.  As a result, the best-fitting mass-richness relation becomes
less steep.  When we change to use a higher concentration $\cvir=7$,
then due to this concentration-mass anti-correlation, the best-fitting
masses are significantly suppressed (which explains the large change
in \mvirt).  Furthermore, this suppression is stronger at the higher
mass end, where the difference between $\cvir=7$ and
Eq.~\eqref{E:powlawcm} is most pronounced.  This trend will tend to
suppress $\gamma$, as is seen in the table.

When we impose corrections from the simulations to the results from
the first fit method, we find that the variation in \mvirt\ and
$\gamma$ when we change the assumed concentration is significantly
reduced.  However, there is still 30 per cent level variation, which
may be ascribed to the fact that the scales that are used in this fit
are quite prone to systematics such as intrinsic alignments and
centroiding errors, which will affect the fits with different assumed
concentrations in different ways.  The simulation corrections can only
correct for the fitting methods' different responses to a theoretical
cluster lensing profile, not for their different responses to
additional systematics that may be present in the data.

When we fit using $\dsr$ from $0.5<R<4$\hmpc, we find smaller
variations in the (uncorrected) amplitude \mvirt\ of the mass-richness relation when
we change the concentration-mass relation, at most $13$ per cent,
which is still problematic since it is close to twice the $1\sigma$
statistical error.  
(However, note that the fit
$\chi^2$ are not sufficiently different to rule out any of these three
models; the lensing data only weakly constrain the concentration.)
The trends in $\gamma$ with $\cvir(\mvir)$ have the same sign as when
fitting using $\dsr$ from $0.1<R<1$\hmpc, but are less pronounced (11
per cent variation, slightly smaller than the $1\sigma$ statistical
error).  Because of the longer range in transverse separation, the
statistical errors on the fit parameters have become smaller, though
we do not fully benefit from this fact due to the systematic
uncertainties.  We also note that for a given concentration-mass
relation, such as Eq.~\eqref{E:powlawcm}, the amplitude \mvirt\ is
increased by 4 per cent relative to the previous results.  This
increase may be due to systematics that decrease the signal on scales
below $0.5$\hmpc, such as intrinsic alignments or centroiding errors.
The fact that $\gamma$ has decreased relative to the $0.1<R<1$\hmpc\
results suggests that the change in masses is more significant at
lower richness than at higher richness.

When we impose corrections from the simulations in
Fig.~\ref{F:showsimfit} to the results of this second fit, we find
that the total range of \mvirt\ and $\gamma$ values is quite small, roughly
7 and 3 per cent respectively.  This finding is encouraging: it suggests
that we may be converging to a result that is more robust to
small-scale systematics.  Since the typical corrected mass from this
fit method is 25 per cent higher than that for the fits using $0.1<R<1$\hmpc, we
conclude that the fits that use those smaller scales may be
significantly influenced by small-scale systematics.

Finally, we consider the results of fits to \ups\ with $R_0=0.5$ using
$0.5<R<4$\hmpc.  First, we see that the statistical errors on fit
parameters are larger than when using \dsr\ for the same scales (50
per cent larger, comparable to or smaller than the errors when using
\dsr\ from $0.1<R<1$\hmpc).  This trend in the errors may seem
inconsistent with the results in the simulations, which suggested
$\sim 30$ per cent increase in mass estimation statistical errors.
However, the 50 per cent increase is for the power-law amplitude that
comes from using 6 mass bins.  On each individual mass bin, the mass
uncertainties increase by 30 per cent when using $\ups$ with
$R_0=\rmin=0.5$ and $\rmax=4$\hmpc\ relative to $\dsr$ with
$\rmin=0.5$ and $\rmax=4$\hmpc.  Thus, the mass increase we see here
for individual mass bins is consistent with that in the simulation.
Second, the variation in the uncorrected \mvirt\ when we change the
concentration-mass relation is 11 per cent, comparable to the
$1\sigma$ errors, though we emphasise again that the variations in
concentration that we have allowed are relatively extreme compared to
what is seen in simulations.  The variation in $\gamma$ is 7 per cent,
more than a factor of two smaller than the statistical error.  The
sense of the change in \mvirt\ when changing $\cvir(\mvir)$ is the
opposite as when fitting to \dsr, as we have seen before in the
simulations.

When we use the simulation results to correct these final fits that
use \ups, we see that the corrections again reduce the spread in the
best-fitting \mvirt\ and $\gamma$ values when we use different
concentration-mass relations.  The residual 4 per cent variation in
both fit parameters is well below the statistical error.  We note that
the typical mass \mvirt\ at richness $\nt=20$ has increased by 10 per
cent relative to the fits using \dsr\ on the same exact scales, even
after the imposition of the correction from simulations in
Fig.~\ref{F:showsimfit}.  We suggest that this change may result from
low-level residual contamination of \dsr\ due to systematics such as
centroiding errors even for $R>\rmin=0.5$\hmpc.  Such contamination
can, as we have shown, bias fits to \dsr\ while not affecting fits to
\ups.  Thus, we adopt our mass normalisation at the pivot richness
$\nt=20$ as $\mvirt/(10^{14}\hmsun) = 1.54 \pm 0.16
\,\,\mathrm{(stat.)} \pm 0.06 \,\,\mathrm{ (sys.)}$, the mean of the
values from the fits to \ups\ with the different concentration-mass
relations.  This systematic error results from an uncertainty of
$0.03$ due to uncertainties in the mass estimation due to both the
assumed and true profile, added in quadrature with the lensing signal
calibration uncertainty of $0.05$.


We now compare these results against the $\mvir(\nt)$ relations
determined in several previous papers.  First, we compare against that
from \cite{2008JCAP...08..006M}, which used these data to fit for a
concentration-mass and mass-richness relation.  Given that the
best-fitting concentration-mass relation in that paper was quite
similar to our Eq.~\eqref{E:powlawcm}, and that the fits in that paper
used \dsr\ from $0.5<R<3$\hmpc, we expect quite similar results to the
results in this paper using $\cvir(\mvir)$ from Eq.~\eqref{E:powlawcm}
and \dsr\ from $0.5<R<4$\hmpc.  
In that paper, we found $\mvirt=1.55$ and
$\gamma=1.14$.  The mass normalisation is quite similar to what we
quote here, because (a) in \cite{2008JCAP...08..006M} the masses were
reduced by approximately 10 per cent due to small-scale systematics
(from the use of \dsr\ rather than \ups), but (b) the lensing signal
amplitude was too high by 6 per cent, as explained above, which raised
the best-fit mass by $1.06^{1.4}$, a 9 per cent difference.  


\cite{2008MNRAS.390.1157R} used the maxBCG cluster lensing data to
estimate a mass-richness relation.  That work used fits to \dsr\ from
$0.5$--$4$\hmpc\ assuming Eq.~\eqref{E:powlawcm} for the
concentration-mass relation, with the same source shape measurements,
shear calibration, and source redshift distribution calibration as in
this paper.  
However, the richness range used in that paper was slightly different,
since it used the entire public catalogue from the minimum $\nt=10$ to
the maximum scaled richness.  Furthermore, the binning into richness
bins within the range that is shared by this work and that one was
different.  Finally, as for \cite{2008JCAP...08..006M}, they
explicitly modelled the halo-halo term using the same halo model
formalism and assumed mass-bias relation.  Their result was a
best-fitting mass-richness power-law with $\mvirt=1.42$ and
$\gamma=1.16$.  Thus, the calibration is 8 per cent lower than
the value we
have adopted here, but this could be attributed to differences in
richness ranges.

Finally, we compare against the fits to the maxBCG catalogue cluster
lensing signal in \cite{2007arXiv0709.1159J}.  The differences in
procedure compared to this paper are numerous.  First, the richness
range is different, because they use a non-public version of the
catalogue that extends down to $\nt=3$.  They fit to \dsr\ using
$0.05\le R\le 30$\hmpc, and allow the halo concentration and the
amplitude of the large-scale structure term to vary.  They also use a
model for BCG centroiding errors based on mock catalogues, and
incorporate this model into their fitting routine to correct for the
tendency of centroid errors to suppress the estimated masses.  They
explicitly include lognormal scatter on the mass-richness relation
(with a strong prior in the fits).  Finally, while they use the same
galaxy shape measurements, they use different photometric redshifts,
which we have shown in \cite{2008MNRAS.386..781M} leads to a
calibration bias in the lensing signal of -15 per cent.  Since we have
found that the fitted masses when assuming an NFW profile scale like
$\ds^{1.4}$, this bias in \ds\ corresponds to a 20 per cent
suppression of the masses.  Thus, while they find $\mvirt=1.2$ and
$\gamma=1.3$ (for a spherical over-density of $180\overline{\rho}$,
which should only differ from our definition by several per cent), we
compare against a corrected value of $\mvirt=1.5$.  This result is
within a few per cent of our value of $\mvirt=1.54$ that we have
adopted here.  Given the different richness range
(which also contributes to the different value of $\gamma$) and the
many other differences in fit procedure, the three per cent discrepancy
is not of concern, and is comparable to our quoted systematic uncertainty.


\section{Conclusions}
\label{S:conclusions}

In this paper, we have assessed the degree to which certain systematic
errors in lensing measurement and methods of mass estimation can bias
weak lensing cluster mass estimates.  In brief, the challenges we
considered included the following.
\begin{itemize}
\item Lensing calibration bias, which leads to changes in the mass
  $\propto \ds^{\eta}$ for $\eta$ typically in the range $1.4$--$2$
  depending on the radial range and fit method used for the parametric
  NFW mass fits (lower for \dsr\ than for \ups), or $\propto \ds$ for
  the non-parametric mass estimates within a fixed physical aperture
  (or a steeper scaling when estimated the mass within some spherical
  over-density radius) using \zetac\
  (Section~\ref{SSS:calibration},~\ref{SSS:resnonpara},
  and~\ref{SSS:ressimpara}).
\item Offsets of the identified BCG from the minimum of the cluster
  potential well (Section~\ref{SSS:offsets}) were incorporated into
  the lensing profiles using a model from mock catalogues presented in
  \cite{2007arXiv0709.1159J}.  This model includes the effects of
  photometric errors in selecting the wrong BCG, and is therefore an
  overly conservative estimate in cases where the BCG can be
  unambiguously identified for all clusters or where X-ray data can
  precisely locate the cluster centre.
\item The effect of differences between an assumed NFW concentration and the true
  NFW concentration were studied using pure NFW lensing signals.
\item Differences in the halo profile relative to a pure NFW profile were
  studied using fits to density profiles from $N$-body simulations.
\item The effects of mass from structures other than the cluster
  itself on the lensing signal were also studied using the signal
  from simulations, since we have not included only the mass that is
  virialized when computing \ds\ in the simulations.
\item Contamination of the source sample by cluster member galaxies,
  intrinsic alignments of those member galaxies, and baryonic effects
  on the halo density profile were considered to be included among the
  previous tests, namely changes in NFW
  concentration (in Section~\ref{SS:theorresults}), changes in
  the inner region of the profile using variations of the $N$-body simulation
  outputs (in Section~\ref{SS:simresults}), and centroid offsets that
  modify the signal only in the inner regions of the cluster.
\end{itemize}

When fitting a parametric model (in our case, the NFW profile) to
\dsr, with fixed concentration, we find that the uncertainties due to
unknown true concentration plus changes in the lensing profile due to
small-scale systematics yield systematic errors that range from a
factor of two in mass (when only using small scales in the fits,
e.g. $0.1$--$1$\hmpc) to tens of per cent (when using $R>0.5$\hmpc) to
several per cent (for $R>2$\hmpc, which yields stable mass estimates
but large statistical errors, and which may not be available for
individual cluster lensing analyses due to limited telescope FOV).
This level of systematic error occurred when allowing a relatively
  broad variation in concentration ($4<\cvir <7$), given the disagreement between
  simulations on the concentration-mass relation at high masses, the
  large lognormal scatter in this relation, and other systematics such
  as baryonic effects discussed in
  Sections~\ref{SS:challenges-theory}, \ref{SS:challenges-obs}, and~\ref{SS:paramodel}.
  When using a narrower range in concentration, the systematic errors
  decreased comparably, but are still unacceptably large relative to
  what is needed for precise cosmological parameter constraints.

The addition of centroiding errors to the list of systematics we
considered led to uniform suppression of the mass estimates of order
tens of per cent (for $\rmin=0.1$\hmpc).  To completely avoid this
suppression while fitting to \dsr\ and ignoring the possibility of
centroiding errors, we found it necessary to restrict the
fits to $\rmin>1$\hmpc.  Generally, the addition of larger
scales, out to $\sim 2$\rvir, is useful in minimising the effects of
small-scale systematics; going beyond that can lead to excessive
contribution from large-scale structure, which will bias the mass
estimates if it is not modelled accurately.  Allowing a variation in
concentration in the fits is another way to reduce systematic error,
at the expense of statistical errors that are increased by 45 per cent, but this scheme is not
helpful when dealing with systematics that have a radial profile that
does not mimic a change in concentration.  \ups\ is still more
reliable at removing the impact of small-scale systematics on the
lensing signal.

The aperture mass statistic \zetac\ led to accurate estimates of
projected masses, provided that either (a) the mass in the outer
annulus was estimated rather than ignored, or (b) the mass in the
outer annulus was ignored, but $R_{o1}\gg R_1$ (i.e. a large range of
transverse separations was included in the first integral in
Eq.~\eqref{E:zetac}).  For many applications, such as the halo mass
function, the quantity of interest
is the 3d virial mass, for which a density profile must be assumed
to do the conversion from the 2d projected mass within $R_1$.  We found that
uncertainty in the true density profile led to tens of per cent
level biases in the 3d virial masses.  The effect of
centroiding errors was to uniformly suppress the aperture masses by
$\sim 10$--$20$ per cent depending on the halo mass, degree of
centroiding errors, and transverse separations used for the analysis;
these biases were then propagated into the 3d enclosed mass
estimates.  The aperture mass-based estimates of the cluster virial
mass were substantially noisier than fits to \dsr\ using the same
range of scales.

Finally, the new statistic we introduce here, \ups, removes the effect
of small scales from the lensing signal, gave superior performance
over \dsr\ when fitting an NFW profile to the cluster lensing signal.
This statement is true not just for the basic tests with pure NFW
profiles and profiles from simulations, but also when including the
effects of centroiding errors.  The increases in statistical error on
the mass can be $\sim 40$ per cent relative to fitting to \dsr\ over
the same scales.  
The residual systematic uncertainties after removal of an overall
offset in the masses is of order several per cent, when fitting from
$0.5<R<4$\hmpc, as demonstrated using SDSS maxBCG data.  The effects of \ups\ in decreasing systematic error
are less dramatic when only small scales ($\le 2$\hmpc) are used for
the mass estimates; however, the residual systematics of order 10 per
cent are still at least a factor of two smaller than when fitting to
\dsr.

These conclusions also apply for individual cluster lensing analyses;
however, we caution that in that case, we expect additional
uncertainties in the true halo profile due to contamination by cluster
member galaxies, the lognormal scatter in concentration at fixed mass,
mergers, substructure, triaxiality, and projection effects
(Section~\ref{SS:challenges-theory} and~\ref{SS:challenges-obs}), so the systematic errors will tend to
be larger than for stacked analyses using the same mass estimation
method.

Next, we will briefly discuss the implications of our findings about
mass estimation methods for several previously-published cluster
lensing studies.  We begin with \cite{2009arXiv0903.1103O}, which
contains an analysis of circularly-averaged cluster lensing data for
thirty individual clusters by comparison with spherical models.  They
begin with direct fitting of the tangential shear profile to
parametric models, including the NFW profile.  These fits allow all
model parameters to vary; for example, the NFW fits have two
parameters, a mass and a concentration, unlike the cases we have
considered here with a fixed concentration.  Consequently, the
estimated masses from the NFW model fits are unlikely to be strongly
biased due to modeling assumptions, since the concentration is not
fixed.  However, they may still have some systematic bias due to the
NFW profile not describing cluster profiles well, due to deviations in
individual cluster profiles due to substructure, mergers, and
triaxiality, and possibly significant biases due to small-scale
systematics such as contamination by cluster member galaxies and
centroiding errors.

\cite{2009arXiv0903.1103O} also use the aperture mass statistic
\zetac\ to estimate \mproj, while neglecting the second term in
Eq.~\eqref{E:zetac} and choosing the outer annulus such that it does not
contain any significant structures.  As we have seen here, the
aperture mass statistic when including both terms properly leads to
quite accurate projected mass estimates, or can yield results that are
accurate at the several per cent level even without the second term
provided that $R_1 \ll R_{o1}$.  Given the scales that are accessible
with the Subaru Suprime-Cam, and the typical cluster redshifts, 
we should compare against the top portion of Table~\ref{T:nfwmap}, the rows
with $(R_1, R_{o1}, R_{o2})=(0.275, 1.1, 2)$ and $(0.5,1.1,2)$\hmpc.
Those results suggest that for the most massive clusters, 
neglect of the second term may cause 15--20 per cent suppression of the
projected masses.  We find that the suppression is reduced to 5--10
per cent for more typical cluster masses of $10^{14}\hmsun$.
Furthermore, as we have already seen, effects that 
suppress the signal in the inner cluster regions, such as centroiding
errors and contamination by cluster member galaxies, can suppress the
aperture masses at the $\sim 10$ per cent level.

\cite{2007MNRAS.379..317H} contains an analysis of cluster weak
lensing data for twenty individual clusters.  This work utilises
parametric mass estimates from the tangential shear distortion
averaged in annuli, fitting to an NFW profile with fixed
concentration-mass relation from $N$-body simulations using
$0.25<R<1.5$\hmpc.  In this case, we can assess systematic uncertainties as
being somewhere between the results for $(\rmin, \rmax) = (0.25, 1)$
and $(0.25, 2)$\hmpc\ on Fig.~\ref{F:showsimfit}.  That figure
suggests that uncertainties due to differences between the assumed and
the true profile lead to $\sim 50$ per cent variations in the
estimated cluster halo masses.  This variation may be manifested as
significant noisiness in the mass estimates for a given true mass, as
well as a mean bias if the true profiles (with the imposition of
systematics such as contamination by cluster member galaxies) differ
from the NFW profile with that assumed concentration-mass relation.
This problem is in addition to other uncertainties in individual
cluster mass estimates noted previously, such as LSS (for which they
explicitly increase their error bars) and triaxiality.

\cite{2007MNRAS.379..317H} also use the aperture mass statistic to
estimate projected masses, \mproj, while estimating the second term in
Eq.~\eqref{E:zetac} due to the outer annulus using the best-fitting
NFW model.  In that case, we note that while
\cite{2007MNRAS.379..317H} do not miss mass by excluding the second
term in the aperture mass calculation, their conversion from
$\mproj(<R_1)$ to virial radii using spherical overdensities that can
define the mass function will be strongly concentration-dependent.
While \cite{2007MNRAS.379..317H} claim that the fact that the masses
from the fits to \dsr\ and from the aperture mass calculation are
consistent shows that their fitting procedure is unbiased, as
discussed in Section~\ref{SSS:resnonpara} this claim is not true.  The
fact that \cite{2006ApJ...640..691V} use the cluster mass estimates
from this work to calibrate their mass function constraints is
therefore of concern, because of the possible biases due to these
systematics in the signal and the large systematic scatter that we
have found.

In summary, we believe that weak lensing is the best observational
technique to robustly estimate cluster virial masses (regardless of
their dynamical state) at the level required for precision
cosmology. Given the small statistical errors of recent cluster
abundance analyses, the cosmological constraints are already dominated
by the systematic precision of the cluster mass determination
\citep{2006ApJ...640..691V}. As we argue in this paper, current
methods are inadequate for this purpose because they rely on the
information from the inner parts of the cluster, which can be
contaminated or modified due to a variety of effects discussed in this
paper, and because they do not use numerical $N$-body simulations to
calibrate their results.  Our results suggest eliminating lensing
information from scales below $R_0$ (for which we suggest the range
$0.2<R_0<0.5$\hmpc\ or about 15-25 per cent of the virial radius, as
determined via an iterative procedure).  Our proposed statistic for
parametric estimates of cluster mass, the ADSD \ups, achieves this by
construction, and is consequently more robust to many different
systematics and to the details of the model to which the data are
fitted, all of which are more problematic in the inner parts of the
cluster.  Use of \ups\ to estimate cluster masses allows systematic
errors to be reduced to the several per cent level, which is up to a
factor of 10 smaller than when fitting to the lensing signal \dsr\
itself, suggesting that for current and future datasets, \ups\ should
be the statistic of choice for parametric mass fitting to cluster weak
lensing data. While we have focused on clusters in this paper, similar
concerns about accurately determining the halo mass would arise also
for smaller halos. For these, the stellar component from the galaxy
(and possibly the associated redistribution of the dark matter) would
modify the mass distribution relative to predictions from pure $N$-body
simulations in the inner parts, suggesting that eliminating the inner
halo information by using \ups\ could lead to more accurate mass
determination of group and galaxy type halos as well.

\section*{Acknowledgments}

We thank the anonymous referee for many useful comments.  
 R.M. was supported for the duration of this project by NASA
through Hubble Fellowship grant \#HST-HF-01199.02-A awarded by the
Space Telescope Science Institute, which is operated by the
Association of Universities for Research in Astronomy, Inc., for NASA, 
under contract NAS 5-26555. U.S. is partly supported 
by the Swiss National Foundation under contract 200021-116696/1, 
Packard Foundation and WCU grant R32-2008-000-10130-0.  T.B. acknowledges support
by a grant of the German Academic Foundation during the initial phase of
this project.  R.E.S. acknowledges support from a Marie Curie
Reintegration grant and the SNF.

Funding for the SDSS and SDSS-II has been provided by the Alfred
P. Sloan Foundation, the Participating Institutions, the National
Science Foundation, the U.S. Department of Energy, the National
Aeronautics and Space Administration, the Japanese Monbukagakusho, the
Max Planck Society, and the Higher Education Funding Council for
England. The SDSS Web Site is {\em http://www.sdss.org/}. 

The SDSS is managed by the Astrophysical Research Consortium for the
Participating Institutions. The Participating Institutions are the
American Museum of Natural History, Astrophysical Institute Potsdam,
University of Basel, University of Cambridge, Case Western Reserve
University, University of Chicago, Drexel University, Fermilab, the
Institute for Advanced Study, the Japan Participation Group, Johns
Hopkins University, the Joint Institute for Nuclear Astrophysics, the
Kavli Institute for Particle Astrophysics and Cosmology, the Korean
Scientist Group, the Chinese Academy of Sciences (LAMOST), Los Alamos
National Laboratory, the Max-Planck-Institute for Astronomy (MPIA),
the Max-Planck-Institute for Astrophysics (MPA), New Mexico State
University, Ohio State University, University of Pittsburgh,
University of Portsmouth, Princeton University, the United States
Naval Observatory, and the University of Washington. 


\bibliographystyle{mn2e}
\bibliography{cosmo,cosmo_preprints}

\begin{thebibliography}{115}
\expandafter\ifx\csname natexlab\endcsname\relax\def\natexlab#1{#1}\fi

\bibitem[{{Abate} {et~al.}(2009)}]{2009ApJ...702..603A}
{Abate} A., {Wittman} D., {Margoniner} V.~E., {Bridle} S.~L., {Gee} P., {Tyson}
  J.~A., {Dell'Antonio} I.~P., 2009, \apj, 702, 603

\bibitem[{{Abazajian} {et~al.}(2003)}]{2003AJ....126.2081A}
{Abazajian} K. et~al. 2003, \aj, 126, 2081

\bibitem[{{Abazajian} {et~al.}(2004)}]{2004AJ....128..502A}
{Abazajian} K. et~al. 2004, \aj, 128, 502

\bibitem[{{Abazajian} {et~al.}(2005)}]{2005AJ....129.1755A}
{Abazajian} K. et~al., 2005, \aj, 129, 1755

\bibitem[{{Abazajian} {et~al.}(2009)}]{2009ApJS..182..543A}
{Abazajian} K.~N. et~al. 2009, \apjs, 182, 543

\bibitem[{{Adelman-McCarthy} {et~al.}(2006)}]{2006ApJS..162...38A}
{Adelman-McCarthy} J.~K. et~al. 2006, \apjs, 162, 38

\bibitem[{{Adelman-McCarthy} {et~al.}(2007)}]{2007ApJS..172..634A}
{Adelman-McCarthy} J.~K. et~al. 2007, \apjs, 172, 634

\bibitem[{{Adelman-McCarthy} {et~al.}(2008)}]{2008ApJS..175..297A}
{Adelman-McCarthy} J.~K. et~al. 2008, \apjs, 175, 297

\bibitem[{{Agustsson} \& {Brainerd}(2006)}]{2006ApJ...644L..25A}
{Agustsson} I., {Brainerd} T.~G., 2006, \apjl, 644, L25

\bibitem[{{Baldauf} {et~al.}(2009)}]{tobiaspaper}
{Baldauf} T., {Smith} R.~E., {Seljak} U., {Mandelbaum} R., 2009,
preprint (arXiv:0911.4973)

\bibitem[{{Barkana} \& {Loeb}(2009)}]{2009arXiv0907.1102B}
{Barkana} R., {Loeb} A., 2009, preprint (arXiv:0907.1102)

\bibitem[{{Becker} {et~al.}(2007)}]{2007ApJ...669..905B}
{Becker} M.~R. et~al. 2007, \apj, 669, 905

\bibitem[{{Bildfell} {et~al.}(2008)}]{2008MNRAS.389.1637B}
{Bildfell} C., {Hoekstra} H., {Babul} A., {Mahdavi} A., 2008, \mnras, 389, 1637

\bibitem[{{Biviano} \& {Girardi}(2003)}]{2003ApJ...585..205B}
{Biviano} A., {Girardi} M., 2003, \apj, 585, 205

\bibitem[{{Blumenthal} {et~al.}(1986){Blumenthal}, {Faber}, {Flores}, \&
  {Primack}}]{1986ApJ...301...27B}
{Blumenthal} G.~R., {Faber} S.~M., {Flores} R., {Primack} J.~R., 1986, \apj,
  301, 27

\bibitem[{{Borgani} \& {Kravtsov}(2009)}]{2009arXiv0906.4370B}
{Borgani} S., {Kravtsov} A., 2009, preprint (arXiv:0906.4370)

\bibitem[{{Bridle} {et~al.}(2009)}]{2008arXiv0802.1214B}
{Bridle} S. et~al. 2009, Annals of Applied Statistics, 3, 6 (arXiv:0802.1214)

\bibitem[{{Broadhurst} {et~al.}(2005)}]{2005ApJ...619L.143B}
{Broadhurst} T., {Takada} M., {Umetsu} K., {Kong} X., {Arimoto} N., {Chiba} M.,
  {Futamase} T., 2005, \apjl, 619, L143

\bibitem[{{Bullock} {et~al.}(2001)}]{2001MNRAS.321..559B}
{Bullock} J.~S., {Kolatt} T.~S., {Sigad} Y., {Somerville} R.~S., {Kravtsov}
  A.~V., {Klypin} A.~A., {Primack} J.~R., {Dekel} A., 2001, \mnras, 321, 559

\bibitem[{{Buote} {et~al.}(2007)}]{2007ApJ...664..123B}
{Buote} D.~A., {Gastaldello} F., {Humphrey} P.~J., {Zappacosta} L., {Bullock}
  J.~S., {Brighenti} F., {Mathews} W.~G., 2007, \apj, 664, 123

\bibitem[{{Clowe} {et~al.}(2006)}]{2006ApJ...648L.109C}
{Clowe} D., {Brada{\v c}} M., {Gonzalez} A.~H., {Markevitch} M., {Randall}
  S.~W., {Jones} C., {Zaritsky} D., 2006, \apjl, 648, L109

\bibitem[{{Clowe} {et~al.}(2004){Clowe}, {De Lucia}, \&
  {King}}]{2004MNRAS.350.1038C}
{Clowe} D., {De Lucia} G., {King} L., 2004, \mnras, 350, 1038

\bibitem[{{Clowe} {et~al.}(1998)}]{1998ApJ...497L..61C}
{Clowe} D., {Luppino} G.~A., {Kaiser} N., {Henry} J.~P., {Gioia} I.~M., 1998,
  \apjl, 497, L61+

\bibitem[{{Corless} \& {King}(2007)}]{2007MNRAS.380..149C}
{Corless} V.~L., {King} L.~J., 2007, \mnras, 380, 149

\bibitem[{{Corless} \& {King}(2009)}]{2009MNRAS.396..315C}
---, 2009, \mnras, 396, 315

\bibitem[{{Davis} {et~al.}(1985)}]{1985ApJ...292..371D}
{Davis} M., {Efstathiou} G., {Frenk} C.~S., {White} S.~D.~M., 1985, \apj, 292,
  371

\bibitem[{{Diaferio} {et~al.}(2005){Diaferio}, {Geller}, \&
  {Rines}}]{2005ApJ...628L..97D}
{Diaferio} A., {Geller} M.~J., {Rines} K.~J., 2005, \apjl, 628, L97

\bibitem[{{Dodelson}(2004)}]{2004PhRvD..70b3008D}
{Dodelson} S., 2004, \prd, 70, 023008

\bibitem[{{Dolag} {et~al.}(2004)}]{2004A&A...416..853D}
{Dolag} K., {Bartelmann} M., {Perrotta} F., {Baccigalupi} C., {Moscardini} L.,
  {Meneghetti} M., {Tormen} G., 2004, \aap, 416, 853

\bibitem[{{Einasto}(1965)}]{Einasto65}
{Einasto} J., 1965, Trudy Inst. Astrofiz. Alma-Ata, 5, 87

\bibitem[{{Eisenstein} {et~al.}(2001)}]{2001AJ....122.2267E}
{Eisenstein} D.~J. et~al. 2001, \aj, 122, 2267

\bibitem[{{Eke} {et~al.}(2001){Eke}, {Navarro}, \&
  {Steinmetz}}]{2001ApJ...554..114E}
{Eke} V.~R., {Navarro} J.~F., {Steinmetz} M., 2001, \apj, 554, 114

\bibitem[{{Fahlman} {et~al.}(1994)}]{1994ApJ...437...56F}
{Fahlman} G., {Kaiser} N., {Squires} G., {Woods} D., 1994, \apj, 437, 56

\bibitem[{{Faltenbacher} {et~al.}(2007)}]{2007ApJ...662L..71F}
{Faltenbacher} A., {Li} C., {Mao} S., {van den Bosch} F.~C., {Yang} X., {Jing}
  Y.~P., {Pasquali} A., {Mo} H.~J., 2007, \apjl, 662, L71

\bibitem[{{Fedeli} {et~al.}(2007)}]{2007A&A...473..715F}
{Fedeli} C., {Bartelmann} M., {Meneghetti} M., {Moscardini} L., 2007, \aap,
  473, 715

\bibitem[{{Finkbeiner} {et~al.}(2004)}]{2004AJ....128.2577F}
{Finkbeiner} D.~P. et~al. 2004, \aj, 128, 2577

\bibitem[{{Fukugita} {et~al.}(1996)}]{1996AJ....111.1748F}
{Fukugita} M., {Ichikawa} T., {Gunn} J.~E., {Doi} M., {Shimasaku} K.,
  {Schneider} D.~P., 1996, \aj, 111, 1748

\bibitem[{{Gao} {et~al.}(2008)}]{2008MNRAS.387..536G}
{Gao} L., {Navarro} J.~F., {Cole} S., {Frenk} C.~S., {White} S.~D.~M.,
  {Springel} V., {Jenkins} A., {Neto} A.~F., 2008, \mnras, 387, 536

\bibitem[{{Gladders} \& {Yee}(2000)}]{2000AJ....120.2148G}
{Gladders} M.~D., {Yee} H. K.~C., 2000, \aj, 120, 2148

\bibitem[{{Gnedin} {et~al.}(2004)}]{2004ApJ...616...16G}
{Gnedin} O.~Y., {Kravtsov} A.~V., {Klypin} A.~A., {Nagai} D., 2004, \apj, 616,
  16

\bibitem[{{Gunn} {et~al.}(1998)}]{1998AJ....116.3040G}
{Gunn} J.~E. et~al. 1998, \aj, 116, 3040

\bibitem[{{Guzik} \& {Seljak}(2002)}]{2002MNRAS.335..311G}
{Guzik} J., {Seljak} U.~., 2002, \mnras, 335, 311

\bibitem[{{Heymans} {et~al.}(2006)}]{2006MNRAS.368.1323H}
{Heymans} C. et~al. 2006, \mnras,
  368, 1323

\bibitem[{{Hirata} \& {Seljak}(2003)}]{2003MNRAS.343..459H}
{Hirata} C., {Seljak} U., 2003, \mnras, 343, 459

\bibitem[{{Hirata} {et~al.}(2007)}]{2007MNRAS.381.1197H}
{Hirata} C.~M., {Mandelbaum} R., {Ishak} M., {Seljak} U., {Nichol} R.,
  {Pimbblet} K.~A., {Ross} N.~P., {Wake} D., 2007, \mnras, 381, 1197

\bibitem[{{Hirata} {et~al.}(2004)}]{2004MNRAS.353..529H}
{Hirata} C.~M. et~al. 2004, \mnras, 353, 529

\bibitem[{{Ho} {et~al.}(2009)}]{2009ApJ...697.1358H}
{Ho} S., {Lin} Y.-T., {Spergel} D., {Hirata} C.~M., 2009, \apj, 697, 1358

\bibitem[{{Hoekstra}(2001)}]{2001A&A...370..743H}
{Hoekstra} H., 2001, \aap, 370, 743

\bibitem[{{Hoekstra}(2003)}]{2003MNRAS.339.1155H}
---, 2003, \mnras, 339, 1155

\bibitem[{{Hoekstra}(2007)}]{2007MNRAS.379..317H}
---, 2007, \mnras, 379, 317

\bibitem[{{Hogg} {et~al.}(2001)}]{2001AJ....122.2129H}
{Hogg} D.~W., {Finkbeiner} D.~P., {Schlegel} D.~J., {Gunn} J.~E., 2001, \aj,
  122, 2129

\bibitem[{{Ivezi{\' c}} {et~al.}(2004)}]{2004AN....325..583I}
{Ivezi{\' c}} {\v Z}. et~al. 2004, Astronomische
  Nachrichten, 325, 583

\bibitem[{{Johnston} {et~al.}(2007)}]{2007arXiv0709.1159J}
{Johnston} D.~E. et~al. 2007, preprint (arXiv:0709.1159)

\bibitem[{{Katgert} {et~al.}(2004){Katgert}, {Biviano}, \&
  {Mazure}}]{2004ApJ...600..657K}
{Katgert} P., {Biviano} A., {Mazure} A., 2004, \apj, 600, 657

\bibitem[{{King} {et~al.}(2001){King}, {Schneider}, \&
  {Springel}}]{2001A&A...378..748K}
{King} L.~J., {Schneider} P., {Springel} V., 2001, \aap, 378, 748

\bibitem[{{Kleinheinrich} {et~al.}(2005)}]{2005A&A...439..513K}
{Kleinheinrich} M. et~al. 2005, \aap, 439, 513

\bibitem[{{Koester} {et~al.}(2007{\natexlab{a}})}]{2007ApJ...660..221K}
{Koester} B.~P. et~al. 2007{\natexlab{a}},
  \apj, 660, 221

\bibitem[{{Koester} {et~al.}(2007{\natexlab{b}})}]{2007ApJ...660..239K}
{Koester} B.~P.et~al. 2007{\natexlab{b}},
  \apj, 660, 239

\bibitem[{{Kravtsov} {et~al.}(2006){Kravtsov}, {Vikhlinin}, \&
  {Nagai}}]{2006ApJ...650..128K}
{Kravtsov} A.~V., {Vikhlinin} A., {Nagai} D., 2006, \apj, 650, 128

\bibitem[{{Levenberg}(1944)}]{levenberg}
{Levenberg} K., 1944, The Quarterly of Applied Mathematics, 2, 164

\bibitem[{{Limousin} {et~al.}(2007)}]{2007ApJ...668..643L}
{Limousin} M. et~al. 2007, \apj, 668, 643

\bibitem[{{Lupton} {et~al.}(2001)}]{2001adass..10..269L}
{Lupton} R.~H., {Gunn} J.~E., {Ivezi{\' c}} Z., {Knapp} G.~R., {Kent} S.,
  {Yasuda} N., 2001, in ASP Conf. Ser. 238: Astronomical Data Analysis Software
  and Systems X, pp. 269--278

\bibitem[{{Mandelbaum} {et~al.}(2005{\natexlab{a}})}]{2005MNRAS.361.1287M}
{Mandelbaum} R. et~al. 2005{\natexlab{a}},
  \mnras, 361, 1287

\bibitem[{{Mandelbaum} {et~al.}(2005{\natexlab{b}})}]{2005MNRAS.362.1451M}
{Mandelbaum} R., {Tasitsiomi} A., {Seljak} U., {Kravtsov} A.~V., {Wechsler}
  R.~H., 2005{\natexlab{b}}, \mnras, 362, 1451

\bibitem[{{Mandelbaum} {et~al.}(2006{\natexlab{a}})}]{2006MNRAS.367..611M}
{Mandelbaum} R., {Hirata} C.~M., {Ishak} M., {Seljak} U., {Brinkmann} J.,
  2006{\natexlab{a}}, \mnras, 367, 611

\bibitem[{{Mandelbaum} {et~al.}(2006{\natexlab{b}})}]{2006MNRAS.372..758M}
{Mandelbaum} R., {Seljak} U., {Cool} R.~J., {Blanton} M., {Hirata} C.~M.,
  {Brinkmann} J., 2006{\natexlab{b}}, \mnras, 372, 758

\bibitem[{{Mandelbaum} {et~al.}(2008{\natexlab{a}}){Mandelbaum}, {Seljak}, \&
  {Hirata}}]{2008JCAP...08..006M}
{Mandelbaum} R., {Seljak} U., {Hirata} C.~M., 2008{\natexlab{a}}, Journal of
  Cosmology and Astro-Particle Physics, 8, 6

\bibitem[{{Mandelbaum} {et~al.}(2008{\natexlab{b}})}]{2008MNRAS.386..781M}
{Mandelbaum} R. et~al. 2008{\natexlab{b}}, \mnras, 386, 781

\bibitem[{{Mantz} {et~al.}(2008)}]{2008MNRAS.387.1179M}
{Mantz} A., {Allen} S.~W., {Ebeling} H., {Rapetti} D., 2008, \mnras, 387, 1179

\bibitem[{{Marian} {et~al.}(2009){Marian}, {Smith}, \&
  {Bernstein}}]{2009ApJ...698L..33M}
{Marian} L., {Smith} R.~E., {Bernstein} G.~M., 2009, \apjl, 698, L33

\bibitem[{{Marquardt}(1963)}]{marquardt}
{Marquardt} D., 1963, SIAM Journal on Applied Mathematics, 11, 431

\bibitem[{{Massey} {et~al.}(2007)}]{2007MNRAS.376...13M}
{Massey} R. et~al. 2007, \mnras, 376, 13

\bibitem[{{Merritt} {et~al.}(2006)}]{2006AJ....132.2685M}
{Merritt} D., {Graham} A.~W., {Moore} B., {Diemand} J., {Terzi{\'c}} B., 2006,
  \aj, 132, 2685

\bibitem[{{Metzler} {et~al.}(2001){Metzler}, {White}, \&
  {Loken}}]{2001ApJ...547..560M}
{Metzler} C.~A., {White} M., {Loken} C., 2001, \apj, 547, 560

\bibitem[{{Metzler} {et~al.}(1999)}]{1999ApJ...520L...9M}
{Metzler} C.~A., {White} M., {Norman} M., {Loken} C., 1999, \apjl, 520, L9

\bibitem[{{Naab} {et~al.}(2007)}]{2007ApJ...658..710N}
{Naab} T., {Johansson} P.~H., {Ostriker} J.~P., {Efstathiou} G., 2007, \apj,
  658, 710

\bibitem[{{Nagai} {et~al.}(2007){Nagai}, {Kravtsov}, \&
  {Vikhlinin}}]{2007ApJ...668....1N}
{Nagai} D., {Kravtsov} A.~V., {Vikhlinin} A., 2007, \apj, 668, 1

\bibitem[{{Navarro} {et~al.}(1996){Navarro}, {Frenk}, \&
  {White}}]{1996ApJ...462..563N}
{Navarro} J.~F., {Frenk} C.~S., {White} S.~D.~M., 1996, \apj, 462, 563+

\bibitem[{{Neto} {et~al.}(2007)}]{2007MNRAS.381.1450N}
{Neto} A.~F. et~al. 2007, \mnras, 381, 1450

\bibitem[{{Okabe} {et~al.}(2009)}]{2009arXiv0903.1103O}
{Okabe} N., {Takada} M., {Umetsu} K., {Futamase} T., {Smith} G.~P.,
2009, preprint (arXiv:0903.1103)

\bibitem[{{Pedersen} \& {Dahle}(2007)}]{2007ApJ...667...26P}
{Pedersen} K., {Dahle} H., 2007, \apj, 667, 26

\bibitem[{{Pier} {et~al.}(2003)}]{2003AJ....125.1559P}
{Pier} J.~R., {Munn} J.~A., {Hindsley} R.~B., {Hennessy} G.~S., {Kent} S.~M.,
  {Lupton} R.~H., {Ivezi{\' c}} {\v Z}., 2003, \aj, 125, 1559

\bibitem[{{Press} {et~al.}(1992)}]{1992nrca.book.....P}
{Press} W.~H., {Teukolsky} S.~A., {Vetterling} W.~T., {Flannery} B.~P., 1992,
  {Numerical recipes in C. The art of scientific computing}. Cambridge:
  University Press, 1992, 2nd ed.

\bibitem[{{Reyes} {et~al.}(2008)}]{2008MNRAS.390.1157R}
{Reyes} R., {Mandelbaum} R., {Hirata} C., {Bahcall} N., {Seljak} U., 2008,
  \mnras, 390, 1157

\bibitem[{{Richards} {et~al.}(2002)}]{2002AJ....123.2945R}
{Richards} G.~T. et~al. 2002, \aj, 123, 2945

\bibitem[{{Rines} \& {Diaferio}(2006)}]{2006AJ....132.1275R}
{Rines} K., {Diaferio} A., 2006, \aj, 132, 1275

\bibitem[{{Rines} {et~al.}(2007){Rines}, {Diaferio}, \&
  {Natarajan}}]{2007ApJ...657..183R}
{Rines} K., {Diaferio} A., {Natarajan} P., 2007, \apj, 657, 183

\bibitem[{{Rines} {et~al.}(2003)}]{2003AJ....126.2152R}
{Rines} K., {Geller} M.~J., {Kurtz} M.~J., {Diaferio} A., 2003, \aj, 126, 2152

\bibitem[{{Rozo} {et~al.}(2010)}]{2010ApJ...708..645R}
{Rozo} E. et~al. 2010, 
  \apj, 708, 645

\bibitem[{{Rudd} {et~al.}(2008){Rudd}, {Zentner}, \&
  {Kravtsov}}]{2008ApJ...672...19R}
{Rudd} D.~H., {Zentner} A.~R., {Kravtsov} A.~V., 2008, \apj, 672, 19

\bibitem[{{Salucci} {et~al.}(2007)}]{2007MNRAS.378...41S}
{Salucci} P., {Lapi} A., {Tonini} C., {Gentile} G., {Yegorova} I., {Klein} U.,
  2007, \mnras, 378, 41

\bibitem[{{Schmidt} {et~al.}(2009)}]{2009PhRvL.103e1301S} {Schmidt}
  F., {Rozo} E., {Dodelson} S., {Hui} L., {Sheldon} E., 2009, PRL,
  103, 051301

\bibitem[{{Schmidt} \& {Allen}(2007)}]{2007MNRAS.379..209S}
{Schmidt} R.~W., {Allen} S.~W., 2007, \mnras, 379, 209

\bibitem[{{Schuecker} {et~al.}(2004)}]{2004A&A...426..387S}
{Schuecker} P., {Finoguenov} A., {Miniati} F., {B{\"o}hringer} H., {Briel}
  U.~G., 2004, \aap, 426, 387

\bibitem[{{Scoccimarro}(1998)}]{1998MNRAS.299.1097S}
{Scoccimarro} R., 1998, \mnras, 299, 1097

\bibitem[{{Seljak}(2000)}]{2000MNRAS.318..203S}
{Seljak} U., 2000, \mnras, 318, 203

\bibitem[{{Sheldon} {et~al.}(2004)}]{2004AJ....127.2544S}
{Sheldon} E.~S. et~al. 2004, \aj, 127, 2544

\bibitem[{{Siverd} {et~al.}(2009){Siverd}, {Ryden}, \&
  {Gaudi}}]{2009arXiv0903.2264S}
{Siverd} R.~J., {Ryden} B.~S., {Gaudi} B.~S., 2009, preprint (arXiv:0903.2264)

\bibitem[{{Smith} {et~al.}(2002)}]{2002AJ....123.2121S}
{Smith} J.~A. et~al. 2002, \aj, 123, 2121

\bibitem[{{Smith}(2009)}]{2009MNRAS.400..851S}
{Smith} R.~E., 2009, \mnras, 400, 851

\bibitem[{{Spergel} {et~al.}(2003)}]{2003ApJS..148..175S}
{Spergel} D.~N. et~al. 2003, \apjs, 148, 175

\bibitem[{{Spergel} {et~al.}(2007)}]{2007ApJS..170..377S}
{Spergel} D.~N. et~al. 2007, \apjs, 170, 377

\bibitem[{{Springel}(2005)}]{2005MNRAS.364.1105S}
{Springel} V., 2005, \mnras, 364, 1105

\bibitem[{{Stoughton} {et~al.}(2002)}]{2002AJ....123..485S}
{Stoughton} C. et~al. 2002, \aj, 123, 485

\bibitem[{{Strauss} {et~al.}(2002)}]{2002AJ....124.1810S}
{Strauss} M.~A. et~al. 2002, \aj, 124, 1810

\bibitem[{{Tucker} {et~al.}(2006)}]{2006AN....327..821T}
{Tucker} D.~L. et~al. 2006,
  Astronomische Nachrichten, 327, 821

\bibitem[{{van den Bosch} {et~al.}(2005)}]{2005MNRAS.361.1203V}
{van den Bosch} F.~C., {Weinmann} S.~M., {Yang} X., {Mo} H.~J., {Li} C., {Jing}
  Y.~P., 2005, \mnras, 361, 1203

\bibitem[{{Vikhlinin} {et~al.}(2006)}]{2006ApJ...640..691V}
{Vikhlinin} A., {Kravtsov} A., {Forman} W., {Jones} C., {Markevitch} M.,
  {Murray} S.~S., {Van Speybroeck} L., 2006, \apj, 640, 691

\bibitem[{{Vikhlinin} {et~al.}(2009)}]{2009ApJ...692.1060V}
{Vikhlinin} A. et~al. 2009, \apj, 692, 1060

\bibitem[{{Wright} \& {Brainerd}(2000)}]{2000ApJ...534...34W}
{Wright} C.~O., {Brainerd} T.~G., 2000, \apj, 534, 34

\bibitem[{{Yang} {et~al.}(2006)}]{2006MNRAS.373.1159Y}
{Yang} X., {Mo} H.~J., {van den Bosch} F.~C., {Jing} Y.~P., {Weinmann} S.~M.,
  {Meneghetti} M., 2006, \mnras, 373, 1159

\bibitem[{{York} {et~al.}(2000)}]{2000AJ....120.1579Y}
{York} D.~G. et~al. 2000, \aj, 120, 1579

\bibitem[{{Zentner} {et~al.}(2008){Zentner}, {Rudd}, \&
  {Hu}}]{2008PhRvD..77d3507Z}
{Zentner} A.~R., {Rudd} D.~H., {Hu} W., 2008, \prd, 77, 043507

\bibitem[{{Zhang} {et~al.}(2008)}]{2008A&A...482..451Z}
{Zhang} Y.-Y., {Finoguenov} A., {B{\"o}hringer} H., {Kneib} J.-P., {Smith}
  G.~P., {Kneissl} R., {Okabe} N., {Dahle} H., 2008, \aap, 482, 451

\bibitem[{{Zhao} {et~al.}(2009)}]{2009ApJ...707..354Z}
{Zhao} D.~H., {Jing} Y.~P., {Mo} H.~J., {B{\"o}rner} G., 2009, \apj, 707, 354

\end{thebibliography}

\end{document}